%% file: main.tex
\documentclass[journal]{IEEEtran}
\usepackage[sort,noadjust]{cite}
\usepackage[cmex10]{amsmath}
\usepackage{array}
\ifCLASSOPTIONcompsoc
  \usepackage[caption=false,font=normalsize,labelfont=sf,textfont=sf]{subfig}
\else
  \usepackage[caption=false,font=footnotesize]{subfig}
\fi
\usepackage{fixltx2e}
\usepackage{url}
\usepackage[latin1]{inputenc}
\usepackage[T1]{fontenc}
\usepackage[english]{babel}
\usepackage{stmaryrd}

\usepackage{cancel}
\usepackage[normalem]{ulem}

\usepackage{Clustered_Joint_Processing}
\usepackage{pgfplots}
\usepackage{subfig,tikz}
\usepackage{color} 
\usepackage{transparent}
\usepackage{import}
\usepackage{amsmath,amssymb,amsthm,mathrsfs,amsfonts,dsfont} 
\newcommand{\cB}{{\mathcal B}}

\newcommand{\cR}{{\mathcal R}}

\newcommand{\cT}{{\mathcal T}}

\newcommand{\cX}{{\mathcal X}}

\newcommand{\bN}{{\mathbb N}}

\newcommand{\bR}{{\mathbb R}}

\newcommand{\bX}{{\bar X}}
\newcommand{\bY}{{\bar Y}}
\newcommand{\bU}{{\bar U}}

\newcommand{\bE}{{\mathbb E}}

\newcommand{\mkv}{-\!\!\!\!\minuso\!\!\!\!-}

\newtheorem{Definition}{Definition}

\begin{document}
%\title{On Layered Transmission in Clustered Cooperative Cellular Architectures}
\title{Distributed Binary Detection with \\[-2mm] Lossy Data Compression  
}
%The Joint Detection and Lossy Compression Problem}


\author{Gil Katz,~\IEEEmembership{Student Member,~IEEE,}
        Pablo Piantanida,~\IEEEmembership{Senior Member,~IEEE,}
        and M\'erouane Debbah,~\IEEEmembership{Fellow,~IEEE}% <-this % stops a space
        \thanks{This research has been supported by the ERC Grant 305123 MORE (Advanced Mathematical Tools for Complex Network Engineering). The material in this paper was presented in part in the 52nd Annual Allerton Conference on Communication, Control and Computing 2014~\cite{Allerton-2014}, and at the 2015 IEEE International Symposium on Information Theory (ISIT)~\cite{7282966}.}
 \thanks{Gil Katz is with Large Networks and Systems Group (LANEAS), CentraleSupélec, 91192 Gif-sur-Yvette, France. Email:   gil.katz@centralesupelec.fr.}
 \thanks{P. Piantanida are with Laboratoire de Signaux et Systèmes (L2S, UMR8506), CentraleSupélec-CNRS-Université Paris-Sud, 91192 Gif-sur-Yvette, France. Email: pablo.piantanida@centralesupelec.fr.}
\thanks{M. Debbah is with Large Networks and Systems Group (LANEAS), CentraleSupélec, 91192 Gif-sur-Yvette, France. Email:  merouane.debbah@centralesupelec.fr. }   
\thanks{Copyright (c) 2014 IEEE. Personal use of this material is permitted. However, permission to use this material for any other purposes must be obtained from the IEEE by sending a request to pubs-permissions@ieee.org.}}

\maketitle

\begin{abstract}
Consider the problem where a statistician in a two-node system receives rate-limited information from a transmitter about marginal observations of a memoryless process generated from two possible distributions. Using its own observations, this receiver is required to first identify the legitimacy of its sender by declaring the joint distribution of the process, and then depending on such authentication it generates the adequate reconstruction of  the observations satisfying an average per-letter distortion. The performance of this setup is investigated through the corresponding \emph{rate-error-distortion region} describing the trade-off between: the communication rate, the error exponent induced by the detection and the distortion incurred by the source reconstruction. In the special case of \emph{testing against independence}, where the alternative hypothesis implies that the sources  are independent,  the optimal rate-error-distortion region is characterized. An application example to binary symmetric sources is given subsequently and the explicit expression for the rate-error-distortion region is provided as well. The case of ``general hypotheses'' is also investigated.  A new achievable rate-error-distortion region is derived based on the use of non-asymptotic \emph{binning},  improving the quality of communicated descriptions. Further improvement of performance in the general case is shown to be possible when the requirement of source reconstruction is relaxed, which stands in contrast to the case of general hypotheses.
\end{abstract}


\IEEEpeerreviewmaketitle

\section{Introduction}

\IEEEPARstart{T}{he} problem of Hypothesis Testing (HT) is very familiar in statistics. Presented with a list of $n$ independent and identically distributed (i.i.d) realizations of some random variable (RV) $X$, a statistician attempts to determine the probability distribution that governs the RV, out of a known list of possible distributions. One popular special case is Binary HT, where only two possible hypotheses exist, usually referred to as $H_0$ and $H_1$. Readers interested in an overview of HT problems can consult~\cite{Lehmann-2005} and references therein.

The problem of Binary HT is formally defined by two types of error probabilities which are commonly referred to as Type I and II probabilities. Denote by $\alpha_n$ the first type error probability given by the probability that $H_1$ is chosen despite $H_0$ being true, while the error probability of the second type $\beta_n$ is defined to be the probability that $H_0$ is chosen while $H_1$ is true. Although the trade-off between the two error events can be investigated in many ways, one common path is to investigate the exponential rate of decay of the error probability of the second type, i.e., $-\lim\limits_{n \to \infty} \frac{1}{n}\log\beta_n^\star(\epsilon)$, while imposing a fixed constraint over the error probability of the first type, i.e., $\alpha_n \leq  \epsilon$ ($\epsilon >0$). \emph{Stein's Lemma}~\cite{Lehmann-2005,Cover-Thomas-1991} provides a closed-form expression for the optimal error exponent in this case,
 \begin{equation}
 -\lim\limits_{n \to \infty} \frac{1}{n}\log\beta_n^\star(\epsilon) = \mathcal{D}(P_0 \| P_1) \ ,
 \end{equation}
where $P_0$ and $P_1$ are the probability distributions implied by hypotheses $H_0$ and $H_1$, respectively, and $\mathcal{D}(\cdot \| \cdot)$ is the \emph{Kullback-Leibler divergence} provided that the measure $P_0$ is \emph{absolutely continuous} resp. to $P_1$, i.e., $P_0 \ll P_1$. It is worth to emphasize that, the optimal exponential rate of decay of the error probability of the second type does not depend asymptotically on the specific constraint over the error probability $\epsilon$ of the first type. 

The situation is substantially more complicated in the case of a distributed detection. If it were possible to transmit all signals to some central location with negligible cost and delay, then the previous theory is in principle applicable. However, due to practical considerations such as energy cost, reliability, survivability, communication bandwidth, compartmentalization, there is never total centralization of information in practice~\cite{4102537}. In this paper, we focus on the problem of distributed hypothesis testing where it is assumed that realizations of different memoryless sources of finite alphabets are observed at different physical locations and thus, nodes are subject to satisfy different types of communication constraints. This work attempts a modest step in the direction of a theory for distributed testing based on lossy data compression which seems to offer a formidable mathematical complexity (see~\cite{720540} and references therein).

\begin{figure}[bt]
	\centering
	\includegraphics[trim=3cm 9cm 7.5cm 4cm, clip=true,width=0.4\textwidth]{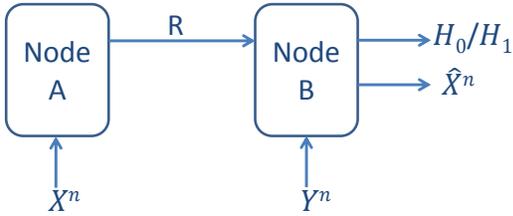}
	\caption{Communication model for joint distributed detection and source reconstruction.}
	\label{fig:Model}
\end{figure}

\subsection{Related Work}
Ahlswede \& Csiszar~\cite{Ahlswede-Csiszar-1986}  and then Han~\cite{Han-1987} investigated the two-node distributed binary HT problem, where only one-sided communication is allowed, with rate $R$ [bits/sample] (see Fig.~\ref{fig:Model} for a representation of a similar system). Both works offer similar approaches to derive achievable rate-exponent rates for this problem, while the results are derived based on somewhat different tools. Although optimality is proven in~\cite{Ahlswede-Csiszar-1986}  for the special case of ``testing against independence'', an optimality result for the general case remains elusive. 

While testing against independence is a particular case that assumes $P_{1,XY} = P_{X}P_{Y}$ and $P_{0,XY} = P_{XY}$, it is important in many scenarios where checking the \emph{relevance} of information being transmitted is of interest. This scenario resembles the known case of transmitting information where side information may be absent \cite{Heegard-Berger-1985,Kaspi-1994}, but is rendered more complex by the fact that even the receiver is unaware of the relevance of the side information. An equivalent setting,  namely vector Gaussian source  coding with decoder side information under mutual information and distortion constraints,  has been investigated in~\cite{5238759}, and benefits of successive refinement for testing against independence are studied in~\cite{4626080}. The problem of testing against independence is approached for the scenario where reciprocal communication is allowed between the two nodes in \cite{Xiang-Kim-2012}. Benefits of a two-way communication system were demonstrated through a coding scheme inspired by the seminal work of Kaspi~\cite{Kaspi-1985}.  

Considering the general HT scenario described in Fig.~\ref{fig:Model}, the problem faced in this paper shares common roots with the seminal works in~\cite{Ahlswede-Csiszar-1986,Han-1987}. Here, however, we are not interested \emph{only} in distributed testing but also in achieving source reconstruction. This also connects to the lossy source coding problem by Heegard \& Berger~\cite{heegard_rate_1985}, where two decoders have to reconstruct the same source based on different side informations and the setup investigated in~\cite{5238759}. Along the line of the technical tools used in the present work, authors in~\cite{Shimokawa-han-amari-1994} suggested  the use of ``binning'' as a possible approach to improve performance of distributed HT by reducing the coding rate. We shall study this approach which, however, brings forth different difficulties,  stemming from the fact that the worth of the side information at the decoder is unknown before a decision is made about the state of the system. That is because reliable decoding of the ``bin index'' is required in presence of side information uncertainty (e.g. similarly to problems under channel uncertainty~\cite{Lapidoth-Narayan-1998}), which is also met and contended with in our present framework. Binning was also shown to be useful in~\cite{Rahman-Wagner-2012}, where a multi-node system composed of several decentralized  encoders that send limited-rate messages to a decoder about their observations was investigated for the case of testing against conditional independence. 

In this work, we consider another dimension of the problem, as represented in Fig.~\ref{fig:Model}. An authentication system prevents the unauthorized injection of messages into a public channel, on which security is inadequate for the needs of its users since it may be threatened with eavesdropping or injection or both~\cite{1055638}. This threat of compromise of the receiver's authentication data is motivated by situations in multiuser networks --such as automatic fault diagnosis-- where the receiver is often the system itself which cannot be treated by conventional cryptography, and which require recourse to new techniques (e.g. image authentication~\cite{5772930, 905982} and Smart Grids~\cite{7035067, 6504815}). Having divided the problem into that of authentication and communication,  decoding of a message at the receiver (node $B$) requires first a reliable identification of the legitimacy of its sender (node A) and then a lossy reconstruction of the underlying feature vector ${X}=(X_1,\dots,X_n)$, with an average per-letter distortion depending on the decision made. In a sense, this problem combines the general distributed HT problem studied in~\cite{Ahlswede-Csiszar-1986} and~\cite{Han-1987} with the classical Heegard \& Berger~\cite{heegard_rate_1985}. 

\subsection{Main Contributions}

The paper is divided into three parts. In the first part, we focus on the case of testing against independence where the alternative hypothesis $H_1$ is a disjoint ``version'' of $H_0$ that leads to $\vct{X}^n$ and $\vct{Y}^n=(Y_1,\dots,Y_n)$ to be independent from each other while sharing the same marginal distributions as under $H_0$. By relying on the techniques introduced in~\cite{Han-1987}, we offer an achievable (single-letter) expression for the tradeoff between the coding rate, the error exponent and the average per-letter distortion, referred  to as \emph{rate-error-distortion region}. In this setting, we simply assume that reconstruction is only attempted when $H_0$ is decided, since no effective side-information is available at the decoder when $H_1$ is the true hypothesis. 

Interestingly, it is shown that the optimal  rate-error-distortion region is attained by using \emph{layered coding}, where the first layer performs HT, and the second layer uses well-known results for source coding with side information at the decoder~\cite{Wyner-Ziv-1976}, while ignoring the information received by node $B$ at the HT stage. This result is quite surprising, as in general there is no reason to believe that such a separation between the two aspects of the problem should be optimal. We explicitly  evaluate the rate-error-distortion region for uniform Binary Sources where a Binary Symmetric Channel (BSC) is assumed between $X$ and $Y$, and plot the resulting tradeoffs between the three quantities of interest.

In the second part, we derive an achievable rate-error-distortion region for the same system, under no specific assumptions on the two hypotheses. To this end, we allow the use of binning not only for source reconstruction but also for the testing purpose. The resulting  rate-error-distortion achievable region is in fact a quadruplet, comprised of the rate of communication, the error exponent for an error of the second type, subject to a maximum probability of error of the first type, and the average distortion corresponding to each hypothesis. The techniques required for this analysis are inspired by previous work on distributed HT~\cite{Han-1987} and recent work~\cite{Kelly-Wagner-2012} on the study of the error exponent for the problem of lossy source coding with side information at the receiver. It should be mentioned here that  although the use of binning for HT was first suggested in~\cite{Shimokawa-han-amari-1994} as a possible approach to improve performance, the benefits of this were never demonstrated. Along this line, Rahman and Wagner~\cite{Rahman-Wagner-2012} show that binning is optimal for HT when under $H_1$ the involved variables  are assumed to be \emph{conditionally independent} given some additional variable, known at the decoder side. While this work played a big part in inspiring a binning approach for HT, it turns out that using $Y$ as the side information available to the receiver does not necessarily improve testing performance, as the exact value of side information is unknown. %In order to further  demonstrate the performance gains attained by the use of binning in HT problems, we introduce the class of ``HT with degraded hypotheses'' which includes a wide class of relevant problems while facilitating calculations.

In the third part of this paper, we concentrate on distributed HT without reconstruction constraints. We show that for the case of two general hypotheses, unlike the case of testing against independence, our previous two-stage coding approach  leads to significant loss in performance. We do so by suggesting a new approach for testing without requiring the decoding of the involved descriptions. This turns out to be superior to the previous one in terms of error exponent, but prevents the decoder of providing a lossy reconstruction of the source. As the performance of the previous approach for general distributed hypotheses testing is lower-bounded by the known result of \cite{Han-1987}, the new approach we introduce may also lead to a significant gain in performance, when compared to this non-binned option.

The rest of this paper is organized as follows. Section~\ref{Sec:Independence} presents the optimal rate-error-distortion region for the case of testing against independence. Optimality is also shown for a specific example of a binary symmetric channel (BSC) between $X$ and $Y$, and numerical results are given. The rate-error-distortion  region for the general HT case is given in Section~\ref{Sec:General}. In Section~\ref{Sec:Focusing},  we offer a different approach for HT only. %The concept of ``degraded hypotheses'' is introduced, and 
The performance of the two previously presented approaches are compared through numerical results. Finally, concluding remarks are given in Section~\ref{Sec:Conclusion}.

\subsection*{Notation and Conventions}
We use upper-case letters to denote random variables (RVs) and lower-case letters to denote realizations of RVs.  Vectors are denoted by boldface letters, with their length as a superscript, emitted when it is clear from the context. Let $\vct{X}_i^j$ denote the vector $\vct{X}$, from position $i$ to position $j$, i.e., $\vct{X}_i^j = (X_i,X_{i+1},\ldots,X_{j-1},X_j)$. $\mathcal{P}(\mathcal{X})$ denotes the set of all possible probability distributions on $\mathcal{X}$, while $p_X \in \mathcal{P}(\mathcal{X})$ is a member of this set. $Q_{\vct{x}^n}$ denotes the empirical distribution, referred to as \emph{the type}, of the vector $\vct{x}^n=(x_1,\dots,x_n)$. $\mathcal{P}_n(\mathcal{X})\subset \mathcal{P}(\mathcal{X})$ denotes the set of all possible atomic probability distributions (or types) on the alphabet $\mathcal{X}$. The set of all vectors $\vct{x}^n \in \mathcal{X}^n$ with a specific type $Q$ is denoted by $\cT(Q) = \cT_{[Q]}$, while the set of all vectors that are $\delta$-typical (in the usual sense, as defined in Appendix~\ref{Apen:typicality}) is denoted by $\cT_{[Q]\delta}^n$. 
Using Csisz\'ar's notation \cite{Csiszar-1998}, we let $H(P_X) = \mathbb{E}\left[-\log p_X(X)\right]$ denote the entropy of a RV distributed according to $p$, and distinguish the binary entropy function by $H_2(x)=-x \log_2 x  -(1-x)\log_2 (1-x)$. $I(X;Y)$ denotes the mutual information between $X$ and $Y$ while assuming that $p_Xp_{Y|X}$ governs the pair, and $\mathcal{D}(P_X\| P^\prime_X)$ the KL divergence between the distributions $p$ and $p'$. All exponents and logarithms in this paper are base $2$, unless stated otherwise. We denote the scalar convolution function by $a\star b \triangleq a(1-b) + b(1-a)$.  Finally, known definitions and properties of typical sequences are given in Appendix~\ref{Apen:typicality}.

%%%%%%%%%%%%%%%%%%%%%%%%%%%%%%%%%%%%%%%%%%%%%%%%%
\section{Testing Against Independence}\label{Sec:Independence}
%%%%%%%%%%%%%%%%%%%%%%%%%%%%%%%%%%%%%%%%%%%%%%%%%

\subsection{Definitions}
In this section, we give a more rigorous formulation of the context depicted in Fig.~\ref{fig:Model} for the case of testing again independence. Let $\mathcal{X}$ and $\mathcal{Y}$ be two finite sets. Nodes A and B observe sequences of random variables $(X_i)_{i\in\mathbb{N}^\star}$ and $(Y_i)_{i\in\bN^\star}$ respectively, which take values on $\mathcal{X}$ and $\mathcal{Y}$, resp. For each $i\in\mathbb{N}^\star$, random samples $(x_i,y_i)$ are distributed according to one of two possible joint distributions:
\begin{equation}\label{eq:TestingAgainstIndependence}
\left\{\begin{aligned}
&H_0: \quad p_0(x,y) = P_{XY}(x,y)\ ,\\
&H_1: \quad p_1(x,y) =  P_{\bX\bY}(x,y) = P_X(x)P_Y(y)\ .
\end{aligned}\right.
\end{equation}
on $\mathcal{X}\times\mathcal{Y}$. Assume that the pairs $(X_i,Y_i)$ are independent across time $i$.

Let $d : \mathcal{X}\times\hat{\mathcal{X}} \to [0\,;d_\text{max}]$ be a finite distortion measure \emph{i.e.}, such that $0\leq d_\text{max} < \infty$. We also denote by $d$ the component-wise mean distortion on $\mathcal{X}^n\times\hat{\mathcal{X}}^n$, \emph{i.e.}, for each $(\vct{x}^n,\vct{\hat{x}}^n)\in\mathcal{X}^n\times\hat{\mathcal{X}}^n$, $d(\vct{x}^n,\vct{\hat{x}}^n) \triangleq \frac1n\,\sum_{i=1}^n d(x_i,\hat{x}_i)$. We assume that node A can send information to node B over an error-free link with rate R bits per source-symbol. Having received the information from node A, node B is then required to make a decision between the two possible hypotheses. After having decided between the two hypotheses, node B attempts to reconstruct the sequence ${X}$, with minimum distortion, for some additive distortion measure, that may depend on the actual probability distribution in place. While recovering the sequence seen by node A under hypothesis $H_1$ may still be possible, it becomes less relevant, as in this case the sequence seen by node B is completely independent and does not constitute as side information. Furthermore, it is very likely that in realistic cases where testing against independence arises, deciding $H_1$ implies that the information seen by node A is irrelevant to node B. Thus, for the case of testing against independence, we assume node B attempts to decode only if it has decided $H_0$. In the general hypotheses case, decoding is attempted under any of the two hypotheses.

\begin{Definition}[Code]\label{Def-cof-independence}
An $(n,R)$-code for testing against independence  in this setup is defined by
\begin{itemize}
\item An encoding function at node A denoted by $f_n : \mathcal{X}^n \to \{1,\dots,\|f_n\|\} $\ ;
\item A decision region $\mathcal{A}_n \subset  \{1,\dots,\|f_n\|\}\times\mathcal{Y}^n$, such that if $(f_n(\vct{x}^n),\vct{y}^n) \in\mathcal{A}_n$ the decoder declares $H_0$ and otherwise  $H_1$\ ;
\item A reconstruction  function at node B 	denoted by $g_n : \{1,\dots,\|f_n\|\}\times \mathcal{Y}^n \to \hat{\mathcal{X}}^n $\ . 
\end{itemize}
\end{Definition}

\begin{Definition}[Rate-exponent-distortion region]\label{def:achievable}
A tuple $(R, E, D,\epsilon)\in\bR_+^4$ is said to be \emph{achievable} if, for any $\delta>0$ and for $n$ large enough, there exists an $(n,R+\delta)$-code $(f_n,\mathcal{A}_n,g_n)$ such that:
\begin{IEEEeqnarray*}{rCl}
n^{-1} \log \|f_n\|&\leq& R\, + \,\delta \ ,\\
\bE_0\big[ d\big(\vct{X}^n,g_n(f_n(\vct{X}^n),\vct{Y}^n)\big) \big]	&\leq& D+\delta \ ,\\
-\dfrac1n\,\log \beta_n(\mathcal{A}_n) &\geq &  E - \delta \ ,\\
\alpha_n(\mathcal{A}_n) & \leq  & \epsilon \ ,
\end{IEEEeqnarray*}
where $\beta_n(\mathcal{A}_n)=\Pr\big(\mathcal{A}_n| XY \sim p_1(x,y) \big)$ and $\alpha_n(\mathcal{A}_n)=\Pr\big(\mathcal{A}_n^c | XY \sim p_0 (x,y) \big)$, and $\bE_0$ denotes that distortion is measured under the condition that node B correctly decides $H_0$. The set of all such achievable tuples is denoted by $\cR^\star$ and is referred to as the \emph{rate-exponent-distortion region}.
\end{Definition}

In~\cite{Ahlswede-Csiszar-1986} and later on in~\cite{Han-1987}, the authors show that when testing against independence, the optimal approach at node $B$ is to apply Stein's Lemma over the common distribution of $\vct{Y}^n$ and the encoded descriptions $f_n(\vct{X}^n)$. More specifically, by optimizing over all decision regions  $\mathcal{A}_n \subset  \{1,\dots,\|f_n\|\}\times\mathcal{Y}^n$, the smallest probability of error of the second type $\beta_n$ asymptotically behaves as: $\beta_n \approx \exp\left(-n E(R)\right)$ with $n$ large enough, for a fixed constraint on the error probability of the first type $\alpha_n \leq \epsilon$, and the exponent $E(R)$ satisfies~\cite[Lemma 1.a]{Ahlswede-Csiszar-1986}:
\begin{equation}\label{Eq:TetaK}
E(R) = \sup\limits_{n\geq 1} E_n(R) \ ,
\end{equation}
where
\begin{equation}
%\begin{aligned}
%& 
E_n(R)=\sup\limits_{f_n} \biggl\{\frac{1}{n} I\left(f_n(\vct{X}^n);\vct{Y}^n\right) \,\Big |\, \log \|f_n \| \leq nR\biggr\} \label{Eq:Equivalence}\ .
%\end{aligned}
\end{equation}
This asymptotic equivalence implies a strong converse property that, much like in the single-node HT setup, the optimal exponential decay of $\beta_n$ is not dependent upon the chosen constraint $0<\epsilon<1$ on the error probability of the first type $\alpha_n$ (e.g. see~\cite{5238759} for a proof based on image sets).  %We now propose a \emph{single-letter} expression for the optimal rate-error-distortion region, as defined in Definition~\ref{def:achievable}. %Exploiting this equivalence the optimal rate-error-distortion region of the system depicted in Fig.~\ref{fig:Model}  can be expressed  through the following \emph{multi-letter characterization}. 

%\begin{Lemma}[Multi-letter characterization~\cite{Ahlswede-Csiszar-1986}] The rate-error-distortion region $\cR^\star$ when testing against independence is described by the set of tuples $(R,E,D)\in\bR_+^3$ satisfying: 
%\begin{IEEEeqnarray}{rCl}\label{Eq:BasicRegion}
%\limsup\limits_{n\rightarrow \infty }\frac{1}{n}\log \|f_n\| & \leq & R \ ,\label{Eq:BasicRegionA} \\
%\liminf\limits_{n\rightarrow \infty } \frac{1}{n} I\left(f_n(\vct{X}^n);\vct{Y}^n\right) & \geq & E \ ,\label{Eq:BasicRegionB}\\
%\limsup\limits_{n\rightarrow \infty } \mathbb{E}_0  \left[d\big(\vct{X}^n, \vct{\hat{X}}^n=g_n(f_n(\vct{X}^n),\vct{Y}^n)\big) \right] &\leq & D \ ,\label{Eq:BasicRegionC}
%\end{IEEEeqnarray}
%for some sequence of encoding and decoding mappings $(f_n,g_n)$. \label{Lemma:Basic}
%\end{Lemma}
%{\color{red} We need to talk about this. I still don't think this is a huge problem, but we need to see what to do with it.}
%
%\begin{Remark}
%Region $\cR^\star$ is closed and convex.
%\end{Remark}

\subsection{Single-Letter Rate-Error-Distortion-Region} \label{sec:single_letter}
We now state the optimal rate-error-distortion region for testing against independence, which provides a single-letter expression for the rate-error-distortion region for testing against independence, defined in  that in Definition~\ref{def:achievable}.
\begin{Prop}[Rate-error-distortion region] \label{Prop:Optimality}
	A tuple $(R,E,D)\in\bR_+^3$  is achievable for the two-node detection and reconstruction problem when testing against independence, as defined in Definition~\ref{def:achievable}, if and only if two random variables $U\in \mathcal{U}$ and $V\in \mathcal{V}$, as well as a reconstruction mapping $g: \mathcal{U}\times \mathcal{V}\times  \mathcal{Y}\to  \hat{\mathcal{X}} $, can be found, such that 
	\begin{IEEEeqnarray}{rCl}
	I(U;X) + I(V;X|UY) &\leq& R \ , \label{Eq:OptRegion-rate} \\
	I(U;Y) &\geq& E \ ,\label{Eq:OptRegion-exponent}\\
	\mathbb{E}_{0}\left[d\big(X,g(UVY)\big) \right]& \leq& D \ ,\label{Eq:OptRegion-distortion}
	\end{IEEEeqnarray}
	with $(U,V)$ being two random variables satisfying  $U \mkv V \mkv X \mkv Y$ form a Markov chain with $(X,Y)\sim p_0(x,y)$, and $\|\mathcal{U}\|  \leq \|\mathcal{X}\| + 2$, $\|\mathcal{V}\|  \leq \|\mathcal{X}\| \|\mathcal{U}\|+ 1$.
\end{Prop}
\begin{IEEEproof}
The proof of Proposition~\ref{Prop:Optimality} is given in Appendix~\ref{Apen:IndependenceAchievability}. \end{IEEEproof}
\begin{Remark}
Observe that on one hand, the expression for the rate can be evaluated as follows:
\begin{equation}\label{Eq:NewRate}
\begin{aligned}
R &\geq  I(U;X) + I(V;X|U) - I(V;Y|U)\\
&= I(U;Y) + \left[I(V;X) - I(V;Y)\right] \ ,
\end{aligned}
\end{equation}
where the final equality stems from the Markov chain formed by the RVs and on the other hand, from the fact that $U \mkv V \mkv X \mkv Y$ form a Markov chain, it is easy to see that 
\begin{equation}
\mathbb{E}_{0}\left[d\big(X,g^\prime (VY)\big) \right] \leq \mathbb{E}_{0}\left[d\big(X,g(UVY)\big) \right]\leq D\ ,
\end{equation}
for some mapping $g^\prime $ and any $g$. Note that the rate can now be seen as comprised of two different parts. The first part of the resulting expression in \eqref{Eq:NewRate} is dedicated to detection since it only affects the error exponent, and is in fact identical to the expression of the error exponent given in \eqref{Eq:OptRegion-exponent} in agreement with previous results \cite{Ahlswede-Csiszar-1986,Han-1987}. The second part of the rate is dedicated only to source reconstruction and therefore, the rate-error-distortion region can be seen as being equivalent to two uncoupled  problems that share a common rate. In the following sections, we will see that this is not the case when general hypotheses are considered.
\end{Remark}
\begin{Remark}\label{remark:CorrectDetectionNotNecessary}
	Note that while the assumption that distortion is only measured in case the detection of hypothesis $H_0$ is convenient, it is not necessary. As we assume that the distortion measure is bound from above, the distortion under the decision $H_0$ (which may or may not be correct) may be expressed as follows:
	\begin{equation}
		\begin{aligned}
		&\mathbb{E}_{0}\left[d\big(X,g(UVY)\big)\right] \\ &= \mathbb{E}_{0}\left[d\big(X,g(UVY)\big)) \text{, ``correct detection''} \right]\\& \qquad \times\Pr\{\text{``correct detection''}\}\\ &+ \mathbb{E}_{0}\left[d\big(X,g(UVY)\big) \text{, ``incorrect detection''} \right]\\& \qquad \times\Pr\{\text{``incorrect detection''}\}\\
		&\leq \mathbb{E}_{0}\left[d\big(X,g(UVY)\big)|H_0 \text{, ``correct detection''} \right] \\&\qquad  + \beta_nd_{\text{max}} \ ,
		\end{aligned}
	\end{equation}
	where $d_{\text{max}}$ is assumed to be that maximal value that the distortion function $d(\cdot,\cdot)$ takes. As $\beta_nd_{\text{max}} \to 0 $ when $n \to \infty$ the relaxation of the assumption that the distortion is only measured under correct detection does not change the optimal rate-error-distortion region. Note that the assumption that estimation is only done under the decision $H_0$ was not relaxed, only the fact that distortion is not measured under incorrect detection.
\end{Remark}

\subsection{Binary Symmetric Source}\label{sec:BSC}
In some cases, the region defined by Proposition~\ref{Prop:Optimality} can be calculated analytically. We present such an example here. Consider the following statistical  model: Let $X \sim \text{Bern}\left(\frac{1}{2}\right)$, and 
\begin{equation}
\begin{aligned}
%&X \sim \text{Bern}\left(\frac{1}{2}\right),\\
&\begin{cases}
H_0: \quad Y = X + Z, \quad Z \sim \text{Bern}(p)\perp X\\
H_1: \quad Y \sim \text{Bern}\left(\frac{1}{2}\right) \perp X  \ ,
\end{cases}
\end{aligned}
\end{equation} 
with $\text{Bern}(p)$ being a \emph{Bernoulli} RV with probability $p$ for being $1$, and $\perp$ signifying that $X$ and $Z$ are independent of each other in the case of hypothesis $0$, and $X$ and $Y$ are independent under the premises of hypothesis $1$. Under both hypotheses, the marginal distributions of both $X$ and $Y$ are equal. Thus, a decision can be reached only through cooperation between the nodes. In the next proposition, the rate-error-distortion region for this problem is characterized  by optimizing over all involved random variables in Proposition~\ref{Prop:Optimality}.

\begin{Prop}[Rate-Error-Distortion region for Binary Symmetric Sources]\label{Prop:BSC}
	The rate-error-distortion region for BSS and testing against independence is given by
%	\begin{IEEEeqnarray}{rCl}
%	R &\geq & 1 - H_2\left(\alpha*\beta*p\right) + \theta\left[H_2\left(\alpha*p\right)-H_2\left(\alpha\right)\right] \label{Eq:RegionBSC-rate}\ ,\\
%	E &\leq & 1 - H_2\left(\alpha*\beta*p\right)\ ,\label{Eq:RegionBSC-exponent}\\
%	D &\geq&  \theta\alpha - \left(1-\theta\right)p \label{Eq:RegionBSC-distortion}\ ,
%	\end{IEEEeqnarray}
	\begin{subequations}
	\begin{align}
		R &\geq  1 - H_2\left(\alpha*\beta*p\right) + \theta\left[H_2\left(\alpha*p\right)-H_2\left(\alpha\right)\right] \label{Eq:RegionBSC-rate}\ ,\\
		E &\leq  1 - H_2\left(\alpha*\beta*p\right)\ ,\label{Eq:RegionBSC-exponent}\\
		D &\geq  \theta\alpha - \left(1-\theta\right)p \label{Eq:RegionBSC-distortion}\ ,
	\end{align}
	\end{subequations}
	
	for any $0 \leq \alpha,\beta \leq \frac{1}{2}$, $0 \leq \theta \leq 1$.
	\end{Prop}
\begin{IEEEproof}
The proof is given in Appendix~\ref{Apen:BSCProof}.
\end{IEEEproof}

\subsection{Numerical Results}
\begin{figure}[t]
	\centering
	\scalebox {0.9} {\subimport{figures/}{error_distortion_independence.tex }}
	\caption{Numerical results of the optimal average distortion as a function of the desired error exponent of the second type, for different amounts of available rate and for $p=0.25$.} 
	\label{Fig:Results}
\end{figure}
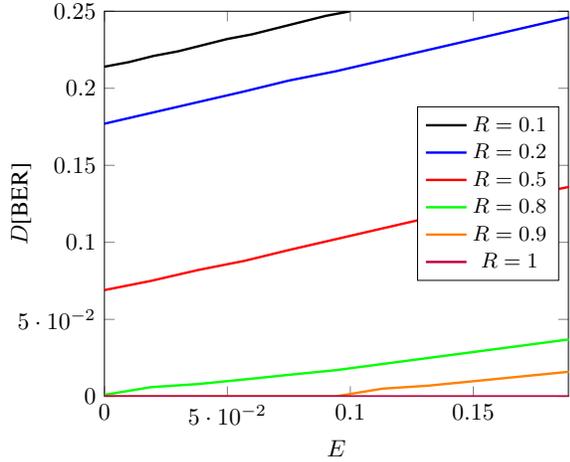

We now present numerical results for the Binary Symmetric Source (BSS) case of testing against independence. Fig.~\ref{Fig:Results} shows six curves, each representing the trade-off between user authentification and source reconstruction, expressed by  the desired error exponent (second type) and the resulting average distortion of the source estimation, for a fixed value of available rate and for $p =0.25$. Unsurprisingly, all curves are non-decreasing, meaning that when the probability of error is exponentially smaller, the amount of rate left for source reconstruction is smaller, resulting in a more crude estimation.

Assuming that both sources $\vct{X}^n$ and $\vct{Y}^n$ are available at a single location, Stein's Lemma yields an error exponent $E_{\max} = I(X;Y) = 1-H_2(p) \approx 0.1887$.  Obviously, this value constitutes an upper bound --uniform over the rate-- on the achievable exponent in the distributed setup presented here. It can be seen that when $R<E_{\max}$, the average distortion reaches its maximal value $D_{\max}=p=0.25$ for some  $E<E_{\max}$. Any exponent bigger than the value for which this happens is unachievable with this rate, since the desired exponent would demand more rate than available. When $R>E_{\max}$, further enlarging the rate allows for better distortion, for the same values of error exponent.

Note especially the curves for the rate values: $R=0.9$ and $R=1$, which comply with $R>H_2(p)$. According to Slepian-Wolf coding (see e.g.~\cite{Cover-Thomas-1991}), this rate is enough to transmit $\vct{x}^n$ to node B without distortion, when no detection is necessary. Indeed, it can be seen that for any choice of error exponent that ensures enough available rate for estimation, zero-distortion is achievable. The curve for $R=1$ is thus almost invisible, as in this case enough rate is available for source reconstruction, for any achievable choice of error exponent.

\section{General Hypothesis Testing}\label{Sec:General}
We now focus on the general case, where both hypotheses can be general distributions of two variables. Note that now, unlike the case of testing against independence, the performance of the system is measured by four quantities, namely the rate, the error exponent and  two distortions, as source reconstruction is attempted under both hypotheses. Nevertheless, distortion is still measured under the assumption that the detection step was completed successfully.  Unlike the case of testing against independence, optimality results for general distributed HT remain elusive. An achievable region~\cite{Han-1987} was inspired by the approach taken for testing against independence. We propose here an achievable region for the general hypothesis testing problem with source reconstruction constraints that makes use of binning for both purposes. The proposed region, while not necessarily optimal in general, aims at improving  on known results for the testing part while also adding the reconstruction of the source.

\subsection{Definitions}

 As before, we suppose that the statistician observes $\vct{Y}^n$ samples directly and can be informed about $\vct{X}^n$ samples indirectly, via an encoding function $f_n:\mathcal{X}^n \to \{1,\dots,\|f_n\|\}$ of rate $n^{-1}\log \|f_n\| \leq R$. The code definition remains the same as in Definition~\ref{Def-cof-independence} with two reconstructions functions $g_{n,i} : \{1,\dots,\|f_n\|\}\times \mathcal{Y}^n \to \hat{\mathcal{X}}_i^n $. For each $i\in\mathbb{N}^\star$, random samples $(x_i,y_i)$ are distributed according to one of two general joint distributions:
\begin{equation} %\label{eq:TestingAgainstIndependence}
\left\{\begin{aligned}
&H_0: \quad p_0(x,y) = P_{XY}(x,y)\ , \\
&H_1: \quad p_1(x,y) = P_{\bX\bY}(x,y)  \ ,
\end{aligned}\right.
\end{equation}
on $\mathcal{X}\times\mathcal{Y}$. Moreover, these samples are independent across time $i=\{1,\dots,n\}$, and we assume throughout this section that $P_X(x) = P_{\bX}(x)$ and $P_Y(y) = P_{\bY}(y)$, $\forall (x,y) \in \mathcal{X}\times\mathcal{Y}$. 

\begin{Definition}[Rate-exponent-distortion region]\label{def:achievable2}
	A tuple $(R, E, D_0,D_1,\epsilon)\in\bR_+^5$ is said to be \emph{achievable} if, for any $\delta>0$, there exists an $(n,R+\delta)$-code $(f_n,\mathcal{A}_n,g_{n,0},g_{n,1})$ such that:
	\begin{equation}
	\begin{aligned}
	n^{-1} \log \|f_n\|&\leq R+\delta \ ,\\
	\bE_i\big[ d_i\big(\vct{X}^n,g_{n,i}(f_n(\vct{X}^n),\vct{Y}^n)\big) \big]	&\leq D_i+\delta 		\ , \text{ $i=0,1$}\\
	-\dfrac1n\,\log \beta_n(\mathcal{A}_n) &\geq   E - \delta \ ,
	\end{aligned}
	\end{equation}
	where $\beta_n(\mathcal{A}_n)=\Pr\big(\mathcal{A}_n| XY \sim p_1(x,y) \big)$ and $\alpha_n(\mathcal{A}_n)=\Pr\big(\mathcal{A}_n^c | XY \sim p_0 (x,y) \big)$, and distortion is measured under the condition that node B correctly detects the correct hypothesis. The set of all such achievable tuples is denoted by $\cR^\star$ and is referred to as the \emph{rate-exponent-distortion region}.
\end{Definition}
\begin{Remark}
	Note the slight abuse of notation in the distortion argument of Definition~\ref{def:achievable2}: The fact that we assume the distortion is measured only in case the detection phase was completed correctly means that for each distortion argument the ``correct'' RVs are assumed to be used. Thus, $\bE_0\big[ d_0\big(\vct{X}^n,g_{n,0}(f_n(\vct{X}^n),\vct{Y}^n)\big) \big]	\leq D_0+\delta$ is the correct expression for the distortion under $H_0$, while $\bE_1\big[ d_1\big(\vct{\bX}^n,g_{n,1}(f_n(\vct{\bX}^n),\vct{\bY}^n)\big) \big]	\leq D_1+\delta$ is the corresponding expression under hypothesis $1$.
\end{Remark}

\subsection{Achievable Rate-Error-Distortion Region}
We now state our main result for the general joint distributed detection and reconstruction problem, which is a new achievable rate-error-distortion region. This region is inspired by the one offered for the special case of testing against independence. In a similar manner to the approach taken in Proposition~\ref{Prop:Optimality}, we derive an achievable region based on the separation of two distinguishable steps, namely user authentication and source reconstruction. The statistician first decodes the description needed to perform testing, and then reconstruct the samples sent by the encoder. However, the decision step requires two phases,  as summarized in the corresponding constraints present in the error exponent of the next proposition.

\begin{Prop}[Achievable rate-error-distortion region]\label{Prop:General}
	A tuple $(R,E,D_0,D_1) \in \mathbb{R}^4_+$, is achievable for the distributed joint detection and reconstruction problem with general hypotheses, if there exists a positive rate $R^\prime$ satisfying: 	
	\begin{equation}
		\begin{aligned}
		R &\geq R^\prime + I\big(X;V_0|UY\big)
		+ I\big(X;V_1|\bU\bY\big) \ , \\
		E &\leq \inf\limits_{Q_X\in\mathcal{P}(\mathcal{X}) }\,\sup\limits_{Q_{U|X}^\star(Q_X) \in\mathcal{P}(\mathcal{U})}\, \inf\limits_{Q_Y\in\mathcal{P}(\mathcal{Y})} \, \\&\inf\limits_{\substack{Q_{UXY}\in\mathcal{P}(\mathcal{U}\times\mathcal{X}\times\mathcal{Y})\\Q_{U|X}=Q_{U|X}^\star}} \Big\{ \min\big[G(Q_{UXY},Q_{X},Q_{Y},R^\prime), \\ 
		&\qquad\qquad \min\limits_{\tilde{U}\tilde{X}\tilde{Y}\in \mathcal{L}(Q^\star_{UX},Q_{UY}^\star)}\mathcal{D}\big(P_{\tilde{U}\tilde{X}\tilde{Y}}\|P_{\bU\bX\bY}\big)\big]\Big\} \,\,\, \label{eq-E-binning}\\
		D_0 &\geq \mathbb{E}_0\left[d_0\big(X,\hat{X}_0(UYV_0) \big)\right] \ ,\\
		D_1 &\geq \mathbb{E}_1\left[d_1\big(\bX,\hat{X}_1(\bU\bY V_1) \big)\right] \ .
		\end{aligned}
	\end{equation}
	Here, $U$ and $\bU$ are auxiliary RVs such that $Q_{U|X}(u|x) = Q_{\bU|\bX}(u|x) \ , \forall (u,x) \in \mathcal{U}\times\mathcal{X}$, $V_0$ and $V_1$ are auxiliary random variables verifying  the Markov chains $U \-- V_0 \-- X \-- Y$ and $\bU \-- V_1 \--\bX \-- \bY$ (along with $U$ and $\bU$ respectively); $\mathcal{L}(Q^\star_{UX},Q_{UY}^\star)$ is the following set of random variables:
	\begin{equation}
	\begin{aligned}
	\mathcal{L}(Q^\star_{UX},Q_{UY}^\star) &= \Big\{P_{\tilde{U}\tilde{X}\tilde{Y}} \in \mathcal{P}(\mathcal{U} \times \mathcal{X} \times \mathcal{Y}) \big|\\&\quad P_{\tilde{U}\tilde{X}}(u,x) = Q^\star_{UX}(u,x),\\ &\quad P_{\tilde{U}\tilde{Y}}(u,y) = Q^\star_{UY}(u,y), \,\forall (u,x,y) \Big\} \ ,\label{eq-set-L}
	\end{aligned}
	\end{equation} 
	\begin{table*}
		\begin{equation}\normalsize
		\begin{aligned}
		G(Q_{UXY},Q_{X},Q_{Y},R^\prime) = 
		&\quad \begin{cases}
		\min\limits_{i = \{0,1\}} \mathcal{D}\big(Q_{UXY}||P_{UXY_i}\big) + \left[R^\prime-I\big(X;U\big) + I\big(Y;U\big)\right]^+ & I\big(U;X\big) > R^\prime \\
		+\infty & \text{else} \ ,
		\end{cases}
		\end{aligned}\label{eq:G}
		\end{equation}
		%\protect\caption{Long equation}
	\end{table*}
	where $Q^\star_{UX},Q^\star_{UY}$ are joint distributions implied by $Q_X$ and the chosen maximizer $Q^\star_{U|X}$,  and the function $G$ appears in \eqref{eq:G}, at the top pf the next page, 
%	\begin{equation}
%	\label{eq:G}
%	\begin{aligned}
%	&G(Q_{UXY},Q_{X},Q_{Y},R^\prime) = \\
%	&\quad \begin{cases}
%	\min\limits_{i = \{0,1\}} \mathcal{D}\big(Q_{UXY}||P_{UXY_i}\big) + \left[R^\prime-I\big(X;U\big) + I\big(Y;U\big)\right]^+ & I\big(U;X\big) > R^\prime \\
%	+\infty & \text{else} \ ,
%	\end{cases}
%	\end{aligned}
%	\end{equation}
	with $P_{UXY_i}$ defined to be $P_{UXY_0} = P_{UXY} = P_{XY}Q_{U|X}$ in the case of hypothesis $0$ and $P_{UXY_1} = P_{\bU\bX\bY} = P_{\bX\bY}Q_{\bU|\bX}$ for hypothesis $1$.
\end{Prop}
%\begin{Remark}
%	The Markovian properties required in Proposition~\ref{Prop:General} can also be written as a single Markov chain of the form $U \mkv (V_0,V_1) \mkv X \mkv (Y_0,Y_1)$. Both of these expressions are equivalent in our case, as $Y_0$ and $Y_1$ are never seen simultaneously, and $V_0$ and $V_1$ never appear  simultaneously. 
%\end{Remark}

\begin{IEEEproof}The proof is relegated to Appendix~\ref{Appen:GeneralAchievability}.\end{IEEEproof} 

We emphasize that  when a binning approach is taken, the expression \eqref{eq-E-binning} for the error exponent $E$ encapsulates the innate tension between two error events: decoding the description and testing based on it. Provided that a good representation $\vct{u}^n$ of the observed samples $\vct{x}^n$ at node A is reliably decoded at node B, the statistician is able to perform detection with a very large probability of success. However, such a good representation  would also induce a very large size for the codebook, which for a given $R$ would cause each bin to be very large in order to satisfy the rate constraint, making likely errors will appear during the  decoding process of the right sequence from the specific bin. On the other hand, when a crude description is chosen, the codebook is smaller and thus so is each bin --if binning is at all necessary. The binning process is therefore not likely to  significantly hurt performance, whereas the retrieved representation is much less valuable for the sake of performing the test because of the crude nature of the description supplied by this representation about samples $\vct{x}^n$. 

In order to ensure the achievability of the error exponent introduced in Proposition~\ref{Prop:General},  we will take a ``worst-case'' approach. The minimization and maximization operators in the expression for $E$ can thus be read as follows: For every possible vector $\vct{x}^n$, the encoder is allowed to choose its strategy of transmission (this is achieved by taking the supremum over $Q^\star_{U|X}$). Having chosen the distribution to generate the codebook, the proposed approach should apply for any type of observed vector $\vct{y}^n$, as well as for any joint type $(\vct{u}^n,\vct{x}^n,\vct{y}^n)$, as long as $Q^\star_{U|X}$ is respected. Much like the case of testing against independence, achievability is proven by dividing the problem into two distinct parts: hypothesis testing and source reconstruction. First, a common message --designed to allow detection-- is communicated from node $A$ to node $B$ and is then used regardless of the probability distribution in effect which is still unknown at this stage. In order to do so, we choose a decoder based on the empirical entropy, similar to the \emph{Empirical Mutual Information} (MMI) decoder used in compound models  (e.g. see~\cite{Lapidoth-Narayan-1998} and references therein). Two private messages are then transposed upon this common message, each intended to be used (together with the common message) under each of the possible hypotheses.  It should be emphasized that dividing the communication in two different phases  may well be a suboptimal choice. However, we will see such a choice introduces significant gains in the error exponent.

\begin{Remark}
	Much like in the case of testing against independence (see Remark~\ref{remark:CorrectDetectionNotNecessary}), the assumption that distortion is only measured when correct detection has occurred is convenient but not necessary for the achievability of the region proposed in Proposition~\ref{Prop:General}. 
\end{Remark}

\section{Focusing on Hypothesis Testing Only}\label{Sec:Focusing}

In this section, we focus on the detection part of the problem only, while still assuming general hypotheses. Although we will show that significant gains can be obtained by introducing  binning as suggested in Proposition~\ref{Prop:General}, we next show that the performance of detection can be further improved  if source reconstruction is not required by the statistician. We start with the following proposition that uses a different approach for testing  without source reconstruction.

\begin{Prop}[Improved error exponent for general hypotheses]\label{Prop:BetterAchievable}
	A pair $(R,E)$ is an achievable rate and exponent pair for general hypothesis testing, without source reconstruction, provided that:
	\begin{equation}
	\begin{aligned}
	E &\leq \sup\limits_{Q_{U|X}^\star\in \mathcal{P}(\mathcal{U})}  \Big\{\min\big\{\hat{G}(Q_{UXY},R)\\&\qquad \min\limits_{\tilde{U}\tilde{X}\tilde{Y}\in \mathcal{L} (Q^\star_{UX}\,,\,Q_{UY}^\star )}\mathcal{D}\big(P_{\tilde{U}\tilde{X}\tilde{Y}}\| P_{\bU\bX\bY}\big)\big\}\Big\}  \ ,\label{eq-error-exponent}
	\end{aligned}
	\end{equation}
	where 
	\begin{equation}
	\hat{G}(Q_{UXY},R) = R  - \left[I\big(X;U\big) -  I\big(U;Y\big)\right]
	\end{equation}
	and the set $\mathcal{L}(Q^\star_{UX},Q_{UY}^\star)$ is defined by \eqref{eq-set-L}. It is worth emphasizing  that $I\big(U;Y\big)$ in  \eqref{eq-error-exponent} is a direct consequence of the choice $Q_{U|X}^\star$ and the distribution implied by $H_0$, $P_{XY}$.
\end{Prop}
\begin{IEEEproof}
	The proof of this proposition is relegated to Appendix~\ref{Apen:BetterAchievability}. 
\end{IEEEproof}

The proof  is very similar to that of Proposition~\ref{Prop:General}. We basically derive the probability of error for a specific triplet of sequences $(\vct{x}^n,\vct{y}^n,\vct{u}^n)$, and then calculate the total probability of error by summing over all possible types and corresponding sequences included within each type. The main difference is that now source reconstruction is not required. Thus, instead of first selecting a sequence from within the bin and only then performing the test, we let node $B$ operate over the entirety of the bin. The chosen strategy consists of going over all sequences within the bin. For each sequence $\vct{u}_i^n$ for $\{1,\dots,2^{nR}\}$, we assume it is the correct one and perform the test by checking the typicality of the pair $(\vct{u}^n_i,\vct{y}^n)$ with relation to the hypothesis $H_0$. If a sequence is found in a bin such that $(\vct{u}^n_i,\vct{y}^n) \in T_{[UY]\delta}^n$, the decoder declares $H_0$. Otherwise, if no such sequence is found it declares $H_1$.

As was the case in Proposition~\ref{Prop:General}, Proposition~\ref{Prop:BetterAchievable} implies that the resulting error exponent is the output of a trade-off between the exponents of the probabilities of two error events. In this case, the trade-off that controls $\beta_n\approx \exp( -n E )$ is between: the probability of erroneous detection while using the right sequence; and the probability of having a different sequence in the bin that is jointly typical with $\vct{y}^n$  and thus would make the decoder declare $H_0$. It turns out, that this trade-off is much preferable to the one offered by Proposition~\ref{Prop:General}, as we can bound the set of sequences that might ``confuse'' the decoder in a manner that is not dependent on the type of $\vct{y}^n$. For instance, the minimizations over $Q_X$, $Q_Y$ and $Q_{UXY}$ (as seen in Proposition~\ref{Prop:General}) are no longer  necessary. This issue has a positive effect on behaviour of the error exponent. Indeed, this new approach takes advantage of the random nature of the binning process. By randomly  allocating sequences into bins we allow for bigger codebooks, which provide better descriptions to the original sequence. As long as the size of the bins are not too large, this does not come at a major price (in terms of the chance of ``confusing" the decoder), and thus improving significantly the result of~\cite{Han-1987} in some cases, as can be seen in the example given subsequently. However, the fact that the original sequence sent by the encoder is not retrieved implies that this strategy is not adapted for the joint problem of detection and source reconstruction. 

\begin{Remark}
Another advantage of this strategy over the one given in Proposition~\ref{Prop:General} is that while knowledge over the probability distribution implied by $P_{\bX\bY}$ is required in order to analyze performance, such  knowledge is not needed in order  to perform the test. This stems from the fact that here, the system only tests if $H_0$ is true or not rather than testing $H_0$ against $H_1$.
\end{Remark}

\subsection{Binary Symmetric Source}
Having proposed two new approaches for distributed testing with general hypotheses, one that allows source reconstruction (Proposition~\ref{Prop:General}) and the other that does not (Proposition~\ref{Prop:BetterAchievable}), it is still not clear whether binning is strictly beneficial for such problems. As was demonstrated in Section~\ref{Sec:Independence}, binning for testing is not necessary  to achieve optimality in the case of testing against independence. One may further argue that as binning introduces additional error events, it is not clear whether or not it would be beneficial at all in the case of general hypotheses. 

In the following, we investigate the benefits of binning through a Binary Symmetric Source (BSS). While it is analytically clear that detection through the strategy offered by Proposition~\ref{Prop:BetterAchievable} is superior to the one offered in Proposition~\ref{Prop:General}, we show that for some specific cases both approaches may result in performance gain relative to non-binning approaches. %This model belongs to the following more general class of problems.
%\begin{Definition}[Degraded hypothesis]
%Let the marginal distribution at each side of a two-node HT problem be equal under both hypotheses. The system can thus be described by the probability distributions $P_X$, $P_{Y_0|X}$ and $P_{Y_1|X}$. Let $H_1$ be \emph{a degraded hypothesis}, with relation to $H_0$, if the channel from $X$ to $Y_1$ is a stochastically degraded version of the channel from $X$ to $Y_0$. 
%	That is, there exists a RV $Y_1$, such that:
%	\begin{equation}\label{Eq:StochasticallyDegraded}
%	P_{Y_1|X}(y|x) = \sum\limits_{\hat{y} \in \mathcal{Y}} P_{Y_0|X}(\hat{y}|x)\hat{P}_{Y_1|Y_0}(y|\hat{y}) \ .
%	\end{equation}
%\end{Definition}
For the sake of simplicity, we consider the following lower bound over the performance, throughout the following numerical analysis~\cite{Han-1987}:
 \begin{equation}
 \min\limits_{\tilde{U}\tilde{X}\tilde{Y}\in \mathcal{L}(Q^\star_{UX},Q_{UY})}\mathcal{D}(P_{\tilde{U}\tilde{X}\tilde{Y}}\|P_{\bU\bX\bY}) \geq \mathcal{D}(P_{UY} \| P_{\bU\bY}) \ .
 \end{equation}
%which simplifies calculations significantly.
 
 Consider the following statistical model: Let $X \sim \text{Bern}\left(\frac{1}{2}\right)$, and
 \begin{equation}
 \begin{aligned}
 &\begin{cases}
 H_0: \quad Y = X + Z_0, \quad Z_0 \sim \text{Bern}(p) \perp X\\
 H_1: \quad Y = X + Z_1, \quad Z_1 \sim \text{Bern}(q)  \perp X\ ,
 \end{cases}
 \end{aligned}
 \end{equation} 
 where $1\geq q > p\geq 0$. Note that while $H_1$ does not imply independence between $X$ and $Y$, the marginal distribution of $Y$ is equal for both hypotheses, making a decision without cooperation impossible. %Furthermore, the channel from $X\to Y_1$ is a stochastically degraded version of the channel from $X\to Y_0$.
 This model was studied first in Wyner-Ziv~\cite{Wyner-Ziv-1976} for source reconstruction. The optimal rate-distortion region  (asymptotic regime) was shown to be
 \begin{equation}
 \begin{cases}
 R(D) = \inf\limits_{\theta,\delta}\left[\theta\left(H_2(p\star\delta) - H_2(\delta)\right)\right] \ ,\\
 D = \theta\delta + (1-\theta)p \ ,
 \end{cases}
 \end{equation}
 where $p$ is the crossover probability between the source $X$ and the side information $Y$, and $p\star \delta$ is the binary convolution of $p$ and $\delta$. The parameters satisfy $0 \leq \theta \leq 1$ and $0 \leq \delta \leq 0.5$. The achievability of this region was shown by using time-sharing between two strategies - in the first the auxiliary RV $U$ is the result of passing $X$ through a Binary Symmetry Channel (BSC) with transition probability $\delta$, while in the second $U$ is degenerate.
 
We now apply Proposition~\ref{Prop:General} to this setup, we choose to consider only distributions in which $Q_X$ is a BSS, and $U$ is the result of passing $X$ through a BSC with crossover probability $\delta$. While this is not necessarily an optimal choice, it can be justified as an optimal approach for the asymptotic regime, at least. To evaluate the resulting error exponent, we need to calculate two values. The first is given by:
\begin{equation}
\inf\limits_{Q_Y} \inf\limits_{\substack{Q_{UXY}\\Q_{U|X}=Q_{U|X}^\star}}G(Q_{UXY},R) \ ,
\end{equation}
as a function of $Q^\star_{U|X}$ (which, under our assumptions, boils down to be a function of $\delta$). This expression encapsulates the error exponent of the event where the wrong sequence is chosen from the bin. The second quantity to calculate is given by:
\begin{equation}\label{eq:SecondCurve}
\min\limits_{\tilde{U}\tilde{X}\tilde{Y}\in \mathcal{L}(Q^\star_{UX},Q_{UY})}\mathcal{D}(P_{\tilde{U}\tilde{X}\tilde{Y}}\|P_{\bU\bX\bY}) \geq \mathcal{D}(P_{UY} \| P_{\bU\bY}) \ ,
\end{equation}
also as a function of $Q^\star_{U|X}$. This expression represents the error exponent of the event where, while using the right sequence, an error occurs during the detection process. Having calculated these two functions, we can pick $Q^\star_{U|X}$ such that the ``minimum'' between the two is ``maximized''.

The results implied by Proposition~\ref{Prop:BetterAchievable} can be calculated in a very similar fashion. Now, the trade-off is between the same curve representing the error while using the correct sequence as was mentioned in \eqref{eq:SecondCurve}, and the curve implied by $\hat{G}$, representing the event of an error caused through the testing of a different sequence.

\subsection{Numerical Results}
\begin{figure}[tb]
	\centering
	\scalebox {0.9} {\subimport{figures/}{BSCGeneral.tex }}
	\caption{Error exponents for both error events in the BSC case with $p=0.1$, $q=0.2$, $R=0.4$, under the strategies implied by Propositions~\ref{Prop:General} and~\ref{Prop:BetterAchievable}. The resulting error exponent for each $\delta$ is the minimum between the two. Performance with a non-binned codebook is represented by a dashed line.} 
\label{Fig:BSCGeneral}\vspace{-0.4cm}
\end{figure}
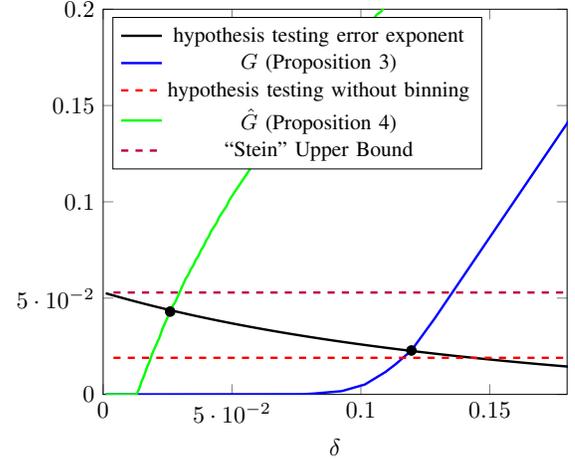
A visualization of the performance achieved by each of the proposed methods for general hypotheses is plotted  in Fig.~\ref{Fig:BSCGeneral}, for the  above discussed statistical model. We choose to consider only distributions in which $Q_X$ is a BSS and $Q_{U|X}^\star$ represents a BSC with transition probability $\delta$, as explained above. The ``hypothesis testing'' curve represents the error exponent of the probability of the event where a mistake is made in detection, when the correct sequence is used from the bin. This curve is relevant for both methods of detection, namely Proposition~\ref{Prop:General} and Proposition~\ref{Prop:BetterAchievable}.

%%%

The interesting tension that exists between the two error events, denoted by either  $G$ (Proposition~\ref{Prop:General}) or $\hat{G}$ (Proposition~\ref{Prop:BetterAchievable}) and an error exponent corresponding to testing, is represented by the worst case between those curves. When $\delta$ is very small, a sequence $\vct{u}^n$ can be found with high probability, such that $\vct{x}^n$ is very well described, and the codebook contains many sequences $\vct{u}^n$. Thus, given the right sequence $\vct{u}^n$, the error event during the test  is not likely, and the error exponent of the event where the test fails is high. However, since the rate of communication is fixed, each bin has to contain many sequences in case $\delta$ is small, increasing the error probability in decoding the right sequence. When $\delta$ grows, the accuracy of the description of $\vct{x}^n$ by $\vct{u}^n$ is lower, making the probability of error of the test, while using the correct sequence, higher. The codebook, however, is smaller, making the task of choosing the right sequence in the bin easier. Note that the error exponent for choosing the sequence from within the bin has a threshold, under which it is zero. This threshold in this case is roughly $\delta \approx 0.08$, which is the value implied by~\cite{Wyner-Ziv-1976} as the minimal value for the binning approach, in the asymptotic regime.

Similarly, the trade-off between the two error events represented by Proposition~\ref{Prop:BetterAchievable} is apparent through the curve  of the error exponent related to the testing errors, along with the ``binning error exponent'' denoted by the curve $\hat{G}$. Now, the additional error event --other than committing an error while using the correct sequence which turns out to be the same as before-- is the event where a different sequence in the bin ``confuses'' the decoder by being jointly typical with $\vct{y}^n$. While this curve is lower bounded by the curve representing $G$ for all cases, it can be seen that in the present case this approach is largely superior. As under both approaches we are allowed to select the strategy $Q_{U|X}^\star$ (in this specific case $\delta$) freely, the optimal approach under each of the propositions would be to choose the corresponding intersection point between the curve representing $G$ or $\hat{G}$ and the curve entitled ``Hypothesis Testing Error Exponent'' in Fig.~\ref{Fig:BSCGeneral}. These two points are marked in Fig.~\ref{Fig:BSCGeneral} with black dots.

In addition, a lower bound can be found in Fig.~\ref{Fig:BSCGeneral}. We emphasize that this bound is not drawn as a function of $\delta$ but rather depicts the best possible performance under the assumptions detailed above, when binning is not performed, as was done in \cite{Han-1987}. Thus, $\delta$ is chosen to be the smallest possible, such that the size of the codebook would not exceed the available rate of communication. A trivial upper bound is also drawn by providing  $\vct{x}^n$ to node $B$ and then applying Stein's Lemma.

\section{Summary and Discussion}\label{Sec:Conclusion}
We studied  the joint problem of distributed detection and lossy compression with side information.  This scenario arises when an authentication system prevents the unauthorized injection of messages into a public channel, assuring the receiver of a message of the legitimacy of its sender. In this setup a user (referred to as node A) is required to communicate a lossy description of a memoryless source to a statistician (referred to as node B) whose task is to verify that the encoding user  is the individual he claims to be and then according to its identity to reconstruct the message based on the adequate distortion measure, much like in \cite{Heegard-Berger-1985,Kaspi-1994}. However, in the setup considered here the receiver is unaware of the value of its information as well, which leads to a two-step approach where first  a decision has to be made about the identity of node A before source reconstruction can take place.

When testing against independence, this two-step approach turns out to be optimal. In this case, detection can be performed optimally as in \cite{Ahlswede-Csiszar-1986}, while source remonstration is performed \`a la Wyner-Ziv~\cite{Wyner-Ziv-1976}, and the two-step approach does not induce performance degradation. An application example to a binary symmetric source was also shown for which the optimal region was explicitly derived, emphasizing an interesting  tension between the error exponent corresponding to the (second type) error probability and the average distortion measure.

When testing with general hypotheses, a similar, albeit more involved, approach produced a new achievable rate-error-distortion region. Here, optimality may be hard to reach, as optimality results stay elusive even in the case where the receiver is aware of the value of the side information (see \cite{Steinberg-Merhav-2004} and references therein). Nevertheless, we showed that the two-step approach, which was optimal in the case of testing against independence, induces in the general case a significant loss in performance. It was shown that when source reconstruction is not required, valuable information for testing  can be compressed much further than in the opposite case,  improving significantly the performance of detection.

Although there are several other threats to authentication systems which require recourse to more sophisticated models and techniques than the ones investigated here, this work attempts a modest step in the direction of a theory for distributed testing based on lossy data compression which seems to offer a formidable mathematical complexity.

%%%%%%%%%%%%%%%%%%%%%% APENDIX %%%%%%%%%%%%%%%%
\appendices
 \section{Typical Sequences and Related Results} \label{Apen:typicality}
   In this appendix we introduce standard notions in information theory, suited for the mathematical developments and proofs needed in this work. The results presented can be easily derived from the standard formulations provided in~\cite{ElGamal-Kim-2011,Csiszar81,Csiszar-1998}. Let $\mathcal{X}$ and $\mathcal{Y} $  be finite alphabets and $(\vct{x}^n,\vct{y}^n)\in\mathcal{X}^n\times \mathcal{Y}^n$.  With $\mathcal{P}(\mathcal{X}\times \mathcal{Y})$ we denote the set of all joint probability distributions on $\mathcal{X}\times\mathcal{Y}$. We define the \emph{$\delta$-typical} sets, with relation to the pmf $p_X \in \mathcal{P}$, as:
    \begin{Definition}[Typical set]\label{Def:TypicalSets}
    	Consider $p\in\mathcal{P}(\mathcal{X})$ and $\delta>0$. We say that $\vct{x}^n\in\mathcal{X}^n$ is $\delta$- typical if $\vct{x}^n\in \mathcal{T}_{[X]\delta}^n$ with:
    	\begin{equation}
	   \begin{aligned}
    	\mathcal{T}_{[X]\delta}^n= & \Big\{\vct{x}^n\in\mathcal{X}^n:\big|Q_{\vct{x}^n}(a)-p_X(a)\big|\leq\delta \ , \\&\quad\quad\quad \forall a\in\mathcal{X}\ \mbox{such that}\ p(a)\neq 0  \Big\},
    	\end{aligned}
    	\label{eq:strong_delta_typ}
    	\end{equation}
    	where $Q_{\vct{x}^n}(a)=n^{-1}N(a|\vct{x}^n)$ is the type of $\vct{x}^n$ and $N(a|\vct{x}^n)$ denotes de number of occurrences of $a\in\mathcal{X}$ in $\vct{x}^n$.  
    \end{Definition}
    
    \begin{Definition}[Joint and conditional typical sets]
    In a similar manner to Definition~\ref{Def:TypicalSets}, given $p_{XY}\in\mathcal{P}\left(\mathcal{X}\times\mathcal{Y}\right)$ we can construct the set of $\delta$-jointly typical sequences as:
    \begin{equation}
    \begin{aligned}
     &\mathcal{T}_{[XY]\delta}^n=\Big\{(\vct{x}^n,\vct{y}^n)\in\mathcal{X}^n\times\mathcal{Y}^n:\\&\Big|Q_{\vct{x}^n\vct{y}^n} (a,b)-p_{XY}(a,b)\Big|\leq\delta, \\
     & \forall (a,b)\in\mathcal{X}\times\mathcal{Y}\ \mbox{ such that } \ p_{Y|X}(b|a)Q_{\vct{x}^n} (a) \neq 0\Big\}\ .
     \label{eq:joint_strong_delta_typ}
    \end{aligned}
    \end{equation}
%  \begin{IEEEeqnarray}{rCl}
%  \mathcal{T}_{[XY]\delta}^n&=&\Big\{(\vct{x}^n,\vct{y}^n)\in\mathcal{X}^n\times\mathcal{Y}^n:\Big|Q_{\vct{x}^n\vct{y}^n} (a,b)-p_{XY}(a,b)\Big|\leq\delta, \nonumber\\
%&& \forall (a,b)\in\mathcal{X}\times\mathcal{Y}\ \mbox{ such that } \ p_{Y|X}(b|a)Q_{\vct{x}^n} (a) \neq 0\Big\}\ .
%  \label{eq:joint_strong_delta_typ}
%  \end{IEEEeqnarray}
   We also define the \emph{conditional} typical sequences. In precise terms, given $\vct{x}^n\in\mathcal{X}^n$ we consider the set:
   \begin{equation}
   \begin{aligned}
      &\mathcal{T}_{[Y|X]\delta}^n(\vct{x}^n) = \Big\{\vct{y}^n\in\mathcal{Y}^n: \\&\Big| Q_{\vct{x}^n\vct{y}^n} (a,b)-p_{Y|X}(b|a) Q_{\vct{x}^n} (a) \Big| \leq \delta,\nonumber\\
      &\forall (a,b)\in\mathcal{X}\times\mathcal{Y}\ \mbox{such that}\ p_{Y|X}(b|a)Q_{\vct{x}^n} (a) \neq 0\Big\}\ .
      \label{eq:cond_strong_delta_typ}
   \end{aligned}
   \end{equation}
%  \begin{IEEEeqnarray}{rCl}
%   \mathcal{T}_{[Y|X]\delta}^n(\vct{x}^n) &=& \Big\{\vct{y}^n\in\mathcal{Y}^n: \Big| Q_{\vct{x}^n\vct{y}^n} (a,b)-p_{Y|X}(b|a) Q_{\vct{x}^n} (a) \Big| \leq \delta,\nonumber\\
% &&\forall (a,b)\in\mathcal{X}\times\mathcal{Y}\ \mbox{such that}\ p_{Y|X}(b|a)Q_{\vct{x}^n} (a) \neq 0\Big\}\ .
%   \label{eq:cond_strong_delta_typ}
%  \end{IEEEeqnarray}
%    Note that alternatively, the same set can be expressed as follows:
%   \begin{equation} 
%   \mathcal{T}_{[Y|X]\delta}^n(\vct{x}^n)=\big\{\vct{y}^n\in\mathcal{Y}^n:(\vct{x}^n,\vct{y}^n)\in\mathcal{T}_{\delta}^n(XY)\big\} 
%   \end{equation}
%   setting $p_{XY}=p_{Y|X} Q_{\vct{x}^n} $. 
   \end{Definition}

   We present the following lemmas without proof.
    
    \begin{lemma}[Properties of typical sets~\cite{Csiszar81}] 
    	The following statements hold:
    	\begin{enumerate}
    		\item  Consider $(\vct{x}^n,\vct{y}^n)\in\mathcal{T}_{[XY]\epsilon}^n$. Then, $\vct{x}^n\in\mathcal{T}_{[X]\epsilon}^n$, $\vct{y}^n\in\mathcal{T}_{[Y]\epsilon}$, $\vct{x}^n\in\mathcal{T}_{X|Y\epsilon}^n(\vct{y}^n)$ and $\vct{y}^n\in\mathcal{T}_{[Y|X]\epsilon}^n(\vct{x}^n)$ \ .
%    		\item Be $\mathcal{T}_{[Y|X]\epsilon}^n(\vct{x}^n)$ with $\vct{x}^n\notin\mathcal{T}^n_{[X]\epsilon}$. Then $\mathcal{T}_{[Y|X]\epsilon}^n(\vct{x}^n)=\varnothing$ \ .
    		\item Be $(\vct{X}^n,\vct{Y}^n)\sim\prod_{t=1}^np_{XY}(x_t,y_t)$. If $\vct{x}^n\in\mathcal{T}_{[X]\epsilon}^n$ we have
    		\begin{equation}
    		\begin{aligned}
    		&\exp\{-n(H(X)+\delta(\epsilon))\} \\& \qquad\leq p_{\vct{X}^n}(\vct{x}^n)\leq \\&\exp\{-n(H(X)-\delta(\epsilon))\}
    		\end{aligned}
    		\end{equation}
    		%\[\exp\{-n(H(X)+\delta(\epsilon))\}\leq p_{\vct{X}^n}(\vct{x}^n)\leq \exp\{-n(H(X)-\delta(\epsilon))\} \]
    		with $\delta(\epsilon)\rightarrow 0$ when $\epsilon\rightarrow 0$. Similarly, if $\vct{y}^n\in\mathcal{T}_{[Y|X]\epsilon}^n(\vct{x}^n)$:
    		\begin{equation}
    		\begin{aligned}
    		&\exp\{-n(H(Y|X)+\delta'(\epsilon))\}\\& \qquad\leq p_{\vct{Y}^n|\vct{X}^n}(\vct{y}^n|\vct{x}^n)\leq \\&\exp\{-n(H(Y|X)-\delta'(\epsilon))\}
    		\end{aligned}
    		\end{equation}
    		%\[\exp\{-n(H(Y|X)+\delta'(\epsilon))\}\leq p_{\vct{Y}^n|\vct{X}^n}(\vct{y}^n|\vct{x}^n)\leq \exp\{-n(H(Y|X)-\delta'(\epsilon))\} \]
    		with $\delta'(\epsilon)\rightarrow 0$ when $\epsilon\rightarrow 0$\ .
    		\label{lemma:cont}
    	\end{enumerate}
    	\label{lemma:useful}
    \end{lemma}
       \begin{IEEEproof}
       	See \cite[Chapter 2.5]{Csiszar81}.
       \end{IEEEproof}
       
    \begin{lemma}[Conditional typicality lemma~\cite{Csiszar81}]
  Consider the product measure $\prod\limits_{t=1}^np_{XY}(x_t,y_t)$,  the following result hold true
    	\begin{IEEEeqnarray*}{rCl}
	\operatorname{Pr}\left\{\mathcal{T}_{[X]\epsilon}^n\right\} &\geq &  1-\mathcal{O}\left(\frac{1}{n\epsilon^2} \right), \\
	\operatorname{Pr}\left\{\mathcal{T}_{[Y|X]\epsilon}^n(\vct{x}^n)|\vct{x}^n\right\}&\geq & 1-\mathcal{O}\left(\frac{1}{n\epsilon^2} \right), \\ \textrm{for every  $\vct{x}^n\in\mathcal{X}^n $}, 
	\end{IEEEeqnarray*}
    	where $(n\epsilon^2) \rightarrow \infty $ when $\epsilon\rightarrow 0$ and $n\rightarrow \infty$.    	\label{lemma:prob_lim}
    \end{lemma}
    \begin{IEEEproof}
    	See \cite[Chapter 2.5]{Csiszar81}.
    \end{IEEEproof}

   \begin{lemma}[Size of typical sets~\cite{Csiszar-1998}]\label{lemma:SizeTypicalSet}
   For any type $Q \in \mathcal{P}_n(\mathcal{X})$
   \begin{equation*}
    |\mathcal{P}_n(\mathcal{X})|^{-1}\exp\big(nH(Q)\big) \leq |\mathcal{T}^n_Q | \leq \exp\big(nH(Q)\big)\ .
   \end{equation*}
   The size of the set of all empirical distributions (or types) of $X$ and of length $n$ can be calculated to be
   \begin{equation*}
   |\mathcal{P}_n(\mathcal{X})| = \binom{n+|\mathcal{X}|-1}{|\mathcal{X}|-1} \leq (n+1)^{|\mathcal{X}|} \ ,
   \end{equation*}
   yielding the following bound
   \begin{equation*}
   (n+1)^{-|\mathcal{X}|}\exp\big(nH(Q)\big) \leq |\mathcal{T}^n_Q| \leq \exp\big(nH(Q)\big) \ .
   \end{equation*}
   \end{lemma}
   
   \begin{lemma}\label{Lemma:SizeDeltaTypicalSet}
   	For every probability measure $P_X \in \mathcal{P}(\mathcal{X})$ and stochastic mapping $W:\mathcal{X}\mapsto \mathcal{P}(\mathcal{Y})$, there exist sequences $(\varepsilon_n)_{n \in \mathbb{N}_+},(\varepsilon^\prime_n)_{n \in \mathbb{N}_+} \to 0$ as  $n \to \infty$ satisfying:
   	\begin{equation}
   	\begin{aligned}
   	&\left|\frac{1}{n} \log |\cT_{[X]_\epsilon}| - H(X)\right| \leq \varepsilon_n \ , \\&  \left|\frac{1}{n} \log |\cT_{[Y|X]_\epsilon}(\vct{x})| - H(Y|X)\right| \leq \varepsilon_n \ , 
   	\end{aligned}
   	\end{equation}
   	for each $\vct{x} \in \cT_{[X]_\epsilon}$ where $\varepsilon_n\equiv \mathcal{O}(n^{-1} \log n) $, and 
   	\begin{equation}
   	\begin{aligned}
   	& P_X^n\big(\cT_{[X]_\epsilon}\big) \geq 1- \varepsilon^\prime_n \ ,\\& W^n\big(\cT_{[Y|X]_\epsilon}(\vct{x})| X^n=\vct{x}\big) \geq 1 - \varepsilon^\prime_n \ ,
   	\end{aligned}
   	\end{equation}
   	for all $\vct{x} \in \mathcal{X}^n$ where $\varepsilon_n^\prime\equiv \mathcal{O}\left(\frac{1}{n\epsilon^2}\right) $, provided that   $n$ is sufficiently large.
   \end{lemma}
   \begin{IEEEproof}
   	Refer to reference~\cite[Lemma 2.13]{Csiszar81}
   \end{IEEEproof}
   
   \begin{lemma}[Set of sequences with small empirical entropy \cite{Kelly-Wagner-2012}]\label{Lemma:EmpiricalEntropy}
For any pair of strings of length $n$, denoted by $(\vct{x}^n,\vct{y}^n)$, let
   \begin{equation*}
   \begin{aligned}
   &\mathcal{S}(\vct{x}^n,\vct{y}^n) = \\&\Big\{(\vct{\tilde{x}}^n,\vct{\tilde{y}}^n)\in \mathcal{X}^n\times \mathcal{Y}^n \, \big | \, H(\vct{\tilde{x}}^n,\vct{\tilde{y}}^n) \leq H(\vct{x}^n,\vct{y}^n)\Big\} \ ,
   \end{aligned}
   \end{equation*}
   with $H(\vct{x}^n,\vct{y}^n)$ being the empirical entropy of the sequences,
   \begin{equation*}
   H(\vct{x}^n,\vct{y}^n) = -\sum\limits_{a \in \mathcal{X}, b \in \mathcal{Y}} Q_{\vct{x}^n \vct{y}^n}(a,b) \log Q_{\vct{x}^n\vct{y}^n}(a,b) \ .
   \end{equation*}
   Then
   \begin{equation*}
   | \mathcal{S}(\vct{x}^n,\vct{y}^n)| \leq (n+1)^{|\mathcal{X}||\mathcal{Y}|} \exp \big[ nH(\vct{x}^n,\vct{y}^n)\big] \ .
   \end{equation*}
   Let 
   \begin{equation*}
   \mathcal{S}(\vct{x}^n |\vct{y}^n) = \Big\{\vct{\tilde{x}}^n\in \mathcal{X}^n\, |\, H(\vct{\tilde{x}}^n|\vct{y}^n) \leq H(\vct{x}^n|\vct{y}^n)\Big \} \ ,
   \end{equation*}
   then
   \begin{equation*}
   |   \mathcal{S}(\vct{x}^n |\vct{y}^n))| \leq (n+1)^{|\mathcal{X}||\mathcal{Y}|}\exp\big[ H(\vct{x}^n|\vct{y}^n)\big] \ .
   \end{equation*}
   \end{lemma}

     \begin{lemma}[Generalized Markov Lemma~\cite{GML_ours_2014}]
     	Let $p_{UXY}\in\mathcal{P}\left(\mathcal{U}\times\mathcal{X}\times\mathcal{Y}\right)$ be a probability measure that satisfies: $U \mkv X \mkv Y$. Consider $(\vct{x},\vct{y})\in\mathcal{T}^n_{[XY]_{\epsilon'}}$ and random vectors $\vct{U}^n$  generated according to:
     	\begin{equation}
     	\begin{aligned}
     	\label{eq:u_dist}
     	& \Pr\left\{\vct{U}^n=\vct{u}\big|{U}^n\in\mathcal{T}_{[U|X]_{\epsilon''}}^n(\vct{x}), \vct{x},\vct{y}\right\}=\\&\frac{\mathds{1}\left\{\vct{u}^n\in\mathcal{T}_{[U|X]_{\epsilon''}}^n(\vct{x})\right\}}{\big|\mathcal{T}_{[U|X]_{\epsilon''}}^n(\vct{x})\big|}\ .
     	\end{aligned}
     	\end{equation}
     	For sufficiently small $\epsilon,\epsilon',\epsilon''>0$,       	
     	\begin{equation}
     	\begin{aligned}
     	&\Pr\left\{\vct{U}^n\notin\mathcal{T}^n_{[U|XY]_{\epsilon}}(\vct{x},\vct{y})\Big|\vct{U}^n\in\mathcal{T}^n_{[U|X]_{\epsilon''}}(\vct{x}),\vct{x},\vct{y}\right\}\\& \quad\equiv \mathcal{O}\left(c^{-n}\right)
     	\end{aligned}
     	\end{equation}
     	holds uniformly on $(\vct{x},\vct{y})\in\mathcal{T}^n_{[XY]_{\epsilon'}}$ where $c>1$.
     	\label{lemma:markov}
     \end{lemma}
     
    \begin{lemma}[Joint Typicality Lemma~\cite{ElGamal-Kim-2011}]\label{Lemma:JointTypicality}
    	Let $(X,Y,Z) \sim p(x,y,z)$ and $\epsilon' < \epsilon$. Then there exist $\delta(\epsilon) >0$ that tends to $0$ as $\epsilon \to 0$ such that the following statements hold:
    	\begin{enumerate}
    		\item If $(\vct{x}^,\vct{y}^n)$ is a pair of arbitrary sequences and $\vct{Z}^n \sim \prod\limits_{i=1}^n p_{Z|X}(z_i|x_i)$ then
    		\begin{equation}
    		\begin{aligned}
    		&\Pr\{(\vct{x}^n,\vct{y}^n,\vct{Z}^n) \in \mathcal{T}^n_{[XYZ]\epsilon}\} \leq \\& \qquad\qquad\exp\{-n(I(Y;Z|X)-\delta(\epsilon))\} \ .
    		\end{aligned}
    		\end{equation}
    		\item If $(\vct{x}^n,\vct{y}^n) \in \cT^n_{[XY]\epsilon'}$ and $\vct{Z}^n \sim \prod\limits_{i=1}^n p_{Z|X}(z_i|x_i)$, then for $n$ sufficiently large 
    		\begin{equation}
    		\begin{aligned}
    		&\Pr\{(\vct{x}^n,\vct{y}^n,\vct{Z}^n) \in \mathcal{T}^n_{[XYZ]\epsilon}\} \leq \\&\qquad\qquad \exp\{-n(I(Y;Z|X)-\delta(\epsilon))\} \ .
        	\end{aligned}
    		\end{equation}    		
    	\end{enumerate}
    \end{lemma}

\section{Proof of Proposition~\ref{Prop:Optimality}}\label{Apen:IndependenceAchievability}
In this appendix, we prove the achievability and converse to Proposition~\ref{Prop:Optimality}.

\subsection*{Achievability proof}
\emph{Codebook generation:} Fix a conditional probability distribution $Q_{VU|XY} = Q_{V|UX}Q_{U|X}P_{XY}$ such that $U \mkv V \mkv X \mkv Y$ form a Markov chain. Let $Q_U(u) = \sum_{x \in \cX}P_X(x)Q_{U|X}(u|x)$ and $Q_{V|U}(v|u) = \sum_{x \in \cX}Q_{V|UX}(v|u,x)$. Let the total available rate of communication $R$ be divided into two, such that the parts are dedicated to $U$ and $V$, which represent the different parts of the message.
Denote the rate dedicated to the transmission of $U$ by $\hat{R}$, while the rate dedicated to the transmission of $V$ is denoted by $R^\prime$. Randomly and independently generate $\exp(n\hat{R})$ sequences $\vct{u}$ through the i.i.d. pmf $Q_U(u)$, with replacement, such that $\vct{u}(s_1) \in \cT_{[U]\delta}$, $\forall s_1$, with $s_1\in[1:\exp(n\hat{R})]$. %from the typical set $\cT_{[U]\delta}^n$ (see Section~\ref{Sec:Tools} for definitions and properties of typical sequences). 
For each codeword $\vct{u}(s_1)$,  randomly and independently generate $\exp(nS_2)$ sequences denoted by $\vct{v}^n(s_1,s_2)$ and indexed with $s_2\in[1: \exp(nS_2)]$ by using the conditional pmf  $Q_{V|U}(\cdot|\vct{u}(s_1))$, with replacement, such that $\vct{v}(s_1,s_2) \in \cT_{[V|U]\delta}(\vct{u}(s_1))$. Divide theses sequences into $\exp[nR^\prime]$ bins, such that each bin contains roughly $\exp[n(S_2-R^\prime)]$ sequences.

\emph{Encoding:} Assuming that the source sequence $\vct{x}^n$ is produced from $X$, look for the first codeword in $U$'s codebook such that $(\vct{u}^n(s_1),\vct{x}^n) \in \cT_{[UX]\delta}^n$. Then, look for the first codeword $\vct{v}^n(s_1,s_2)$ s.t. $(\vct{v}^n(s_1,s_2),\vct{x}^n) \in \cT_{[VX|U]\delta}^n(\vct{u}(s_1))$. Let $b$ be the bin of $\vct{v}^n(s_1,s_2)$. Send the message $f(\vct{x}^n) = (s_1, b)$ to node B.

\emph{Decoding:} Given $\vct{u}(s_1),b$ and $\vct{y}^n$, the decoder first checks if $(\vct{u}^n(s_1),\vct{y}^n) \in \cT_{[UY]\delta}^n$. If so, it declares $H_0$ and otherwise it declares $H_1$. If the decoder decides $H_0$, it then attempts to decode the message (with average distortion $D$) based on $\vct{v}(s_1,s_2)$. This codeword is first recovered by looking in the bin $b$ for the unique codeword such that $\vct{v}^n(s_1,s_2) \in \cT_{[V|UY]\delta}^n(\vct{u}(s_1),\vct{y}^n)$. Then, a per-letter function $g(\cdot)$ is applied  over the entire available information ($U,V$ and $Y$) in order to produce a reconstruction of the source.

\emph{Error events and constraints:} We start with the HT part, and the relation between the expression $I(U;X)$ and the achievable error exponent. Denoting by $\mathcal{B}_0$ the event ``an error occurred during encoding'' (of the HT part $U$), we expand its probability as $\text{Pr}(\mathcal{B}_0) \leq \text{Pr}(\mathcal{B}_1)+\text{Pr}(\mathcal{B}_2)$ with:
\begin{equation}
\begin{aligned}
&\text{Pr}(\mathcal{B}_1) \triangleq \text{Pr} \{\vct{X}^n \notin \cT_{[X]\delta}^n\}\ , \\
&\text{Pr}(\mathcal{B}_2) \triangleq \text{Pr}\{\nexists s_1 \text{ s.t. } (\vct{u}(s_1),\vct{X}^n) \in \cT_{[UX]\delta}^n |\\&\qquad\qquad\qquad\qquad\qquad\qquad  \vct{X}^n \in T_{[X]\delta}^n\}\ ,
\end{aligned} 
\end{equation}
being the probabilities that the source $X$ produces a non-typical sequence, and that (for a typical source sequence) the codebook doesn't contain an appropriate codeword, respectively. From the Asymptotic Equipartition Property (AEP), $\text{Pr}(\mathcal{B}_1) \leq \eta_n^{(1)} \underset{n \to \infty}{\longrightarrow} 0$. As for $\text{Pr}(\mathcal{B}_2)$:
\begin{subequations}
	\begin{align}
	&\text{Pr}(\mathcal{B}_2) = \\
	&=\left(\text{Pr}\{(\vct{U}^n,\vct{X}^n)\notin \cT_{[UX]\delta}^n|\right.\\
	&\qquad\qquad\qquad\left.\vct{U}^n \in \cT_{[U]\delta}^n, \vct{X}^n \in \mathcal{T}_{[X]\delta}^n \} \right)^{\exp({n\hat{R}})}\nonumber\\
	&= \left(1-\text{Pr}\{(\vct{U}^n,\vct{X}^n)\in \cT_{[UX]\delta}^n|\right.\\
	&\qquad\qquad\qquad\left.\vct{U}^n \in \cT_{[U]\delta}^n, \vct{X}^n \in \cT_{[X]\delta}^n \} \right)^{\exp({n\hat{R}})}\nonumber\\
	&\leq \exp[-\exp({n\hat{R}})\text{Pr}\{(\vct{U}^n,\vct{X}^n)\in \cT_{[UX]\delta}^n|\nonumber\\
	&\qquad\qquad\qquad\vct{U}^n \in \cT_{[U]\delta}^n, \vct{X}^n \in T_{[X]\delta}^n \}]\label{Eq:ProofUniEncError}\\
	&\leq \exp[{-\exp({n\hat{R}})\exp({-n\big(I(U;X)+\eta_n^{(2)})\big)}}]\\
	&= \exp\{{-\exp[{-n\big(I(U;X)-\hat{R}+\eta_n^{(2)}\big)}}]\} \ .
	\end{align}
\end{subequations}
Here, inequality \eqref{Eq:ProofUniEncError} is due to the inequality $(1-a)^n \leq \exp({an})$~\cite{Cover-Thomas-1991}. Since $\eta_n^{(2)}\underset{n \to \infty}{\longrightarrow}0$, $\text{Pr}(\cB_2) \to 0$ if $\hat{R}>I(U;X)$. 

\emph{Analysis of $\alpha_n$:} Calculating the probability of error of the first type, $\alpha_n$, boils down to the following:
\begin{subequations}
	\begin{align}
	\alpha_n &= \Pr(H_1|XY \sim P_{XY})\\
	&\leq \Pr(\mathcal{B}_0) \\
	&+ \Pr\{(\vct{U}^n,\vct{Y}^n)\notin \cT_{[UY]\delta}^n|\nonumber\\
	&\qquad\vct{U}^n \in \cT_{[U]\delta}^n, (\vct{U}^n,\vct{X}^n) \in \cT_{[UX]\delta}^n ,XY \sim P_{XY}  \} \label{Eq:UniAlpha1}\\
	& \leq \Pr(\mathcal{B}_0) + \eta^{(3)} \ . \label{Eq:UniAlpha2}
	\end{align}
\end{subequations}
Here, \eqref{Eq:UniAlpha1} is due to the fact that when calculating the probability of error of Type I, we may assume that the true distribution controlling the RVs is the one implied by hypothesis $0$. \eqref{Eq:UniAlpha2}, with $\eta^{(3)} \to 0$, is due to the Generalized Markov Lemma (see Lemma~\ref{lemma:markov} in Appendix~\ref{Apen:typicality}). Thus, it may be concluded that $\alpha_n \to 0$ when $n \to \infty$, and thus $\alpha_n \leq \epsilon$ for any constraint $\epsilon >0$ and $n$ large enough.

\emph{Analysis of $\beta_n$:} Next, we look at the achievable error exponent of Type II with the proposed encoding scheme. For the sake of this analysis, we can assume that hypothesis $H_1$ is the correct one. We will follow steps similar to the ones used in \cite{Xiang-Kim-2012}:
\begin{equation}
\beta_n = \Pr(H_0|XY \sim P_XP_Y) = \Pr(\cB_1^c) \Pr(\cB_0^c|\cB_1^c) \ , %
\end{equation}
where the event $\cB_0$ is defined by
\begin{equation}
\cB_0 = \{(\vct{U}(s_1),\vct{Y}) \notin \mathcal{T}^n_{[UY]\delta'}\}
\end{equation}
to be the event that the \emph{chosen} sequence $\vct{U}(s_1)$ is not jointly typical with the observed sequence $\vct{Y}$.  The term $\Pr(\cB_1^c)$ goes to $1$ when $n$ is large thanks to Lemma~\ref{lemma:prob_lim}. Note that this also means that with large probability an index $s_1$ is chosen out of the codebook, thanks to the Covering Lemma~\cite{ElGamal-Kim-2011} and the fact that we enforce $\hat{R} \geq I(U;X)$. Moreover, note that even if a sequence $s_1$ cannot be found in the codebook, this \emph{does not constitute a problem} for the analysis of $\beta_n$, as in this case the decoder declares $H_1$.

The term $\Pr(\cB_0^c|\cB_1^c)$ can be developed through the Joint Typicality Lemma (see Lemma~\ref{Lemma:JointTypicality}) as follows:
\begin{equation}
\Pr(\cB_0^c|\cB_1^c) \leq \exp\{-n(I(U;Y)  - \epsilon(\delta'))\} \ ,
\end{equation}
for $n$ large enough and with $\epsilon(\delta') \to 0$ as $\delta' \to 0$. Thus $-\lim\frac{1}{n} \log\beta_n \geq I(U;Y)  - \epsilon(\delta')$ when $n$ is large enough, which completes the achievability of the desired error exponent.

\emph{Analysis of the Estimation Phase:} Finally, we show that given a (correct) decision $H_0$, the RV $V$ can be used to decode $\vct{X}^n$ with the desired distortion: Denoting by $\cB_3$ the event ``an error occurred during encoding or decoding'' (of $V$), we expand its probability as follows $\text{Pr}(\cB_3) \leq \text{Pr}(\cB_4)+\text{Pr}(\cB_5)$, with $\text{Pr}(\cB_4)$ being the probability that no codeword $\vct{v}(s_1,s_2)$ could be found in the codebook for the given sequence $\vct{x}^n$ and the chosen codeword $\vct{u}(s_1)$, and $\text{Pr}(\cB_5)$ being the probability that a different codeword in the same bin $b$ is compatible with $\vct{y}^n$ and $\vct{u}(s_1)$. 
\begin{equation}\label{eq:CalcP2}
\begin{aligned}
&\text{Pr}(\cB_4)\\
& \triangleq \text{Pr}\{\nexists s_2 \text{ s.t. } (\vct{v}^n(s_1,s_2),\vct{x}^n)\in \cT_{[VX|U]\delta}^n(\vct{u}^n(s_1))\}\\
&= \left[\text{Pr}\{(\vct{V}^n,\vct{X}^n) \notin \cT_{[VX|U]\delta}^n(\vct{u}(s_1))|\right.\\
&\left.\qquad{V}^n \in \cT_{[V|U]\delta}^n(\vct{u}(s_1)), \vct{X}^n \in \cT_{[X]\delta}^n(\vct{u}(s_1))\}\right]^{\exp({nS_2})}\\
%&\leq 2^{-2^{nS_2}\text{Pr}\{(V^n,X^n) \in T_{\delta}^n(VX|u(s_1))|V^n \in T_{\delta}^n(V|u(s_1)), X^n \in T_{\delta}^n(X|u(s_1)))\}}\\
&\leq \exp\Big\{{-\exp({nS_2})\exp[{-n\left(I(V;X|U)+\eta_n^{(6)}\right)}]}\Big\} \\
& = \exp\Big\{{-\exp[{-n\left(I(V;X|U)-S_2+\eta_n^{(6)}\right)}]}\Big\} \ .
\end{aligned}
\end{equation}
Thus, $\text{Pr}(\cB_4) \underset{n \to \infty}{\longrightarrow} 0$ if $S_2> I(V;X|U)$. Finally,
\begin{equation}
\begin{aligned}
&\text{Pr}(\cB_5) \triangleq \text{Pr} \{\exists s_2^\prime \in b \\&\qquad\qquad\text{ s.t. } \vct{v}^n(s_1,s_2^\prime) \in  \cT_{[V|UY]\delta}^n(\vct{u}^n(s_1),\vct{y}^n)\} \ ,
\end{aligned}
\end{equation}
with $b$ being the bin sent to node B.
\begin{equation}
\begin{aligned}
&\text{Pr}(\cB_5) \leq \exp[{n\left(S_2-R^\prime+\epsilon\right)}]\\&\qquad\times\text{Pr} \{\vct{V}^n \in \cT_{[V|UY]\delta}^n(\vct{u}^n(s_1),\vct{y}^n)|\\&\qquad\qquad\qquad\qquad{V}^n \in \cT_{[V|U]\delta}^n(\vct{u}^n(s_1))\}\\
&\leq \exp[{n\left(S_2-R^\prime+\epsilon\right)}]\\&\qquad\times\exp[{-n\left(I(V;Y|U)+\eta_n^{(7)}\right)}]\\
& = \exp\Big[{-n\left(I(V;Y|U)-(S_2-R^\prime)+\eta_n^{(7)}-\epsilon\right)} \Big] \ .
\end{aligned}
\end{equation}
Thus, $\text{Pr}(\cB_5) \underset{n \to \infty}{\longrightarrow} 0$ if $S_2-R^\prime < I(V;Y|U)$, or equivalently
\begin{subequations}
	\begin{align}
	R^\prime &> S_2 - I(V;Y|U) > I(V;X|U) - I(V;Y|U)\\
	& = I(V;XY|U) - I(V;Y|U) = I(V;X|UY) \ , \label{Eq:UniEndOfIndependence}
	\end{align}
\end{subequations}
where equality \eqref{Eq:UniEndOfIndependence} stems from the Markov chain $U \mkv V \mkv X \mkv Y$. Thus, since the total rate $R$ is composed of $\hat{R}$ and $R^\prime$, we conclude that our scheme is achievable if $R> I(U;X) + I(V;X|UY)$.\footnote{We explicitly ignored an additional error event, which is that $\vct{y}^n$ is not typical. The probability of this event goes to $0$ much like $\Pr(\mathcal{B}_1)$, thanks to the AEP.} 
%Notice that we do not need to check the case that for the true $s_2$, $({v}^n(s_1,s_2),{y}^n) \notin T_{\delta}^n(VY|{u}^n(s_1))$. That is because we only decode under the decision $H_0$, and we are interested in the distortion only when this decision is correct. This means that $({x}^n,{y}^n) \in T_{\delta}^n(XY)$. Together with the coding process and the Markov chain $U \-- V \-- X \-- Y$, the typicality of $({v}^n,{y}^n)$ is assured with high probability through the Markov lemma and basic properties of typical sequences, as summarized in Appendix~\ref{Apen:typicality}. 

We now know that our scheme allows the decoding of $\vct{v}^n$ with high probability when the rate is large enough. It remains to be shown that $V$ (together with $U$ and $Y$, which are also known at node B) is enough to recover $X$ with average distortion $D$. We choose a (possibly suboptimal) decoder, that decodes $x_i$ only from $(u_i,v_i)$ and $y_i$:
\begin{subequations}
	\begin{align}
	d\big(&\vct{x}^n,\vct{\hat{x}}^n(\vct{u}^n,\vct{v}^n,\vct{y}^n)\big) = \frac{1}{n} \sum\limits_{i=1}^n d\big(x_i, \hat{x}(u_i,v_i,y_i)\big) \\
	&{=}  \sum\limits_{\forall(x,u,v,y)} 
	d\big(x,\hat{x}(u,v,y)\big)Q_{\vct{x}^n\vct{u}^n\vct{v}^n\vct{y}^n} (x,u,v,y) \label{Eq:ProofUniDistortion1}\\
	&{\leq} \mathbb{E}_0 \left[d(X,\hat{X}(UVY))\right]\\
	&+ \sum\limits_{\forall(x,u,v,y)} 
	\left|Q_{\vct{x}^n\vct{u}^n\vct{v}^n\vct{y}^n} (x,u,v,y)-p(x,u,v,y)\right| \label{Eq:ProofUniDistortion2}\\
	&{\leq} \mathbb{E}_0 \left[d(X,\hat{X}(UVY)) \right] + d_{\max} |\mathcal{X}| |\mathcal{U}| |\mathcal{V}| 
	|\mathcal{Y}|\delta_n \ , \label{Eq:ProofUniDistortion3}
	\end{align}
\end{subequations}
where the summation in \eqref{Eq:ProofUniDistortion1} and \eqref{Eq:ProofUniDistortion2} is over all the possible letters in the respective alphabets of the RVs $(x,u,v,y) \in \mathcal{X}\times\mathcal{U}\times\mathcal{V}\times\mathcal{Y}$ and inequality \eqref{Eq:ProofUniDistortion3} holds since $(\vct{x}^n,\vct{u}^n,\vct{v}^n,\vct{y}^n) \in \cT_{[XUVY]\delta}^n$. Since $\delta_n\underset{n \to \infty}{\longrightarrow} 0$, the condition $D > \mathbb{E}_0 \left[d\big(X,\hat{X}(UVY)\big) \right]$ is sufficient to achieve distortion $D+\epsilon$ at node B. This concludes the proof of achievability. 

\subsection*{Converse proof}
For this part of the proof we use the \emph{multi-letter} converse result in \cite{Ahlswede-Csiszar-1986}, which states that when no estimation is required, 
\begin{IEEEeqnarray}{rCl}\label{Eq:BasicRegion}
	\limsup\limits_{n\rightarrow \infty }\frac{1}{n}\log \|f_n\| & \leq & R \ ,\label{Eq:BasicRegionA} \\
	\liminf\limits_{n\rightarrow \infty } \frac{1}{n} I\left(f_n(\vct{X}^n);\vct{Y}^n\right) & \geq & E \ .\label{Eq:BasicRegionB}
\end{IEEEeqnarray}
Clearly, this rate-error relationship cannot be beat when \emph{an additional constraint} (in this case, relating to the estimation requirement) is put on the system.

Denote by $W=f(\vct{X}^n)$ the message sent from node A to node B. The rate can be bounded as follows:
\begin{subequations}
	\begin{align}
	nR &\geq I(W;\vct{X}^n)\\
	&=I(W;\vct{X}^n,\vct{Y}^n) = I(W;\vct{Y}^n) + I(W;\vct{X}^n|\vct{Y}^n) \label{Eq:ProofUniConverse1}\\
	&= \sum\limits_{i=1}^nI(W,\vct{Y}^{i-1};Y_i)+ \sum\limits_{i=1}^n I(W;X_i|\vct{Y}^n,\vct{X}^{i-1})\\
	&= \sum\limits_{i=1}^nI(W,\vct{Y}^{i-1};Y_i) \nonumber\\&+ \sum\limits_{i=1}^n I(W;X_i|Y_i,\vct{Y}_{i+1}^n,\vct{Y}^{i-1},\vct{X}^{i-1})\\
	&= \sum\limits_{i=1}^n\left[I(W,\vct{Y}^{i-1};Y_i)\right.\nonumber\\&\quad\left.+ I(W,\vct{Y}_{i+1}^n,\vct{Y}^{i-1},\vct{X}^{i-1};X_i|Y_i)\right] \label{Eq:ProofUniConverse2}\\
	&=\sum\limits_{i=1}^n\left[I(W,\vct{Y}^{i-1};Y_i) + I(W,\vct{Y}^{i-1};X_i|Y_i)\right.\nonumber\\&\quad\left.+ I(\vct{Y}_{i+1}^n,\vct{X}^{i-1};X_i|Y_i,\vct{Y}^{i-1},W)\right] \\
	&= \sum\limits_{i=1}^n \left[ I(W,\vct{Y}^{i-1};Y_i,X_i)\right.\nonumber\\&\quad\left.+ I(\vct{Y}_{i+1}^n,\vct{X}^{i-1};X_i|Y_i,\vct{Y}^{i-1},W)\right]\\
	&= \sum\limits_{i=1}^n \left[ I(W,\vct{Y}^{i-1};X_i)\right.\nonumber\\&\quad\left.+ I(\vct{Y}_{i+1}^n,\vct{X}^{i-1};X_i|Y_i,\vct{Y}^{i-1},W)\right] \ . \label{Eq:ProofUniConverse3}
	\end{align}
\end{subequations}
Here, \eqref{Eq:ProofUniConverse1} and \eqref{Eq:ProofUniConverse3}are due to the Markov chains $W \-- \vct{X}^n \-- \vct{Y}^n$ and $W \-- X_i \-- Y_i$, respectively. \eqref{Eq:ProofUniConverse2} stems from the fact that both sources $X$ and $Y$ are assumed to be jointly i.i.d. Defining $U_i \triangleq (W,\vct{Y}^{i-1})$ and $V_i \triangleq (U_i,\vct{Y}_{i+1}^n,\vct{X}^{i-1})$ the Markov chain $U_i \-- V_i \-- X_i \-- Y_i$ is satisfied since the sources $X$ and $Y$ are assumed to be jointly i.i.d, and the bound over the rate becomes
\begin{equation}
\begin{aligned}
R &\geq \frac{1}{n} \sum\limits_{i=1}^n \left[I(U_i;X_i) + I(V_i;X_i|U_i,Y_i)\right]\\&= I(U;X)+I(V;X|UY) \ ,
\end{aligned}
\end{equation}
with $U$ and $V$ defined through time-sharing as is subsequently shown in \eqref{Eq:TimeSharing}. 

The error exponent can now be expressed as follows:
\begin{equation}
\begin{aligned}
I(W;\vct{Y}^n) &= \sum\limits_{i=1}^n I(W,\vct{Y}^{i-1};Y_i)\\&= \sum\limits_{i=1}^n I(U_i;Y_i) = nI(U;Y) \ ,
\end{aligned}
\end{equation}
with the same definition of $U_i$. Thus, the converse over the error exponent is proved with equality.

Finally, the distortion at node B can be bounded as follows. Define the function $\hat{X}_i$ as the $i$-th coordinate of the estimate in node B:
\begin{equation}
\hat{X}_i(U_i,V_i,Y_i) \triangleq g_i (W,\vct{Y}^{i-1},Y_i,\vct{Y}_{i+1}^n) \ .
\end{equation}
The component-wise mean distortion thus verifies
\begin{equation}\label{Eq:TimeSharing}
\begin{aligned}
D+\epsilon &\geq \mathbb{E}_0 \left[d\big(\vct{X}^n,g(W,\vct{Y}^n)\big)\right]
%& =& \frac{1}{n}\sum\limits_{i=1}^n \mathbb{E}_0\left[d(X_i,\hat{X}_i(U_i,V_i,Y_i))\right]\\
\\&= \frac{1}{n}\sum\limits_{i=1}^n \mathbb{E}_0 \left[d\big(X_Q,\hat{X}_Q(U_Q,V_Q,Y_Q)\big)|Q=i\right]\\
& = \mathbb{E}_0\left[d\big(X_Q,\hat{X}_Q(U_Q,V_Q,Y_Q)\big)\right]
\\&=\mathbb{E}_0 \left[d\big(X,\hat{X}(U,V,Y)\big)\right] \ .
\end{aligned}
\end{equation}
For the sake of this calculation, we use the fact that any $U_i$ and $V_i$, as they were defined for this converse, contain the entire message $W$, as well as the past and future of $Y$. This concludes the converse proof in Proposition~\ref{Prop:Optimality}.

\subsection*{Cardinality bounds}
It remains to establish that the cardinality bounds specified by the conditions in Proposition~\ref{Prop:Optimality} do not affect the minimization. Toward that end we invoke the support lemma \cite[p. 310]{Csiszar81} in order to deduce that $\mathcal{U}$ must have $\| \mathcal{X}\| -1 $ letters in order to ensure preservation of $p(x|u)$ plus three more to preserve the constraints on $D$, $I(U;X)$ and $I(U;Y)$, so $\| \mathcal{U}\| \leq \| \mathcal{X}\| + 2$ suffices. Similarly, $ \mathcal{V}$ must have $\| \mathcal{X}\|\| \mathcal{U}\|  -1 $ letters  in order to ensure preservation of $p(x,u|v)$ plus two more to preserve $D$, and $I(X;V|UY)$. Thus, it suffices to have $\| \mathcal{V}\| \leq \| \mathcal{X}\|\| \mathcal{U}\| + 1$.

\section{Proof of Proposition~\ref{Prop:BSC}}\label{Apen:BSCProof}
\subsection*{Achievability proof}
In order to achieve the region proposed in Theorem~\ref{Prop:BSC}, choose $V$ as the output of a Binary Symmetric Channel (BSC) with cross-over probability $\alpha$ when the input is $X$. Choose $U$ as the output of another BSC, with cross-over probability $\beta$, when the input is $V$:
\begin{equation}\label{Eq:UandVBSC}
\begin{aligned}
V &= X + W_1, \quad W_1 \sim \text{Bern}\left(\alpha\right) \ ,\\
U &= V + W_2, \quad W_2 \sim \text{Bern}\left(\beta\right) \ .
\end{aligned}
\end{equation}
Calculating the expression for the error exponent, $U$ and $Y$ can be thought of as connected through a BSC with cross-over probability $\alpha\star \beta\star p$, which yields:
\begin{equation}
I(U;Y) = H(U) - H(U|Y) = 1-H_2(\alpha\star \beta\star p) \ .
\end{equation}
This complies with the expression proposed in Theorem~\ref{Prop:BSC}. The relation between the second term in the expression for the rate and the amount of distortion expected can be calculated through the following two steps, inspired by the approach taken in \cite{Wyner-Ziv-1976}, for the case of source estimation with side information, jointly distributed according to a BSC (without uncertainty in the probability distribution of the sources):

a) Setting $\hat{X} = g(Y,V) = V$, we have $\mathbb{E}_0\left[d(X,\hat{X})\right] = \alpha$. Note that all expectations henceforth are taken over the distribution imposed by $H_0$, and under the assumption that the decision $H_0$ was correct. $Y$ and $V$ can be thought of as being connected through a BSC with cross-over probability $\alpha\star p$. Thus \eqref{Eq:NewRate} results in
\begin{equation}
\begin{aligned}
R_a &= I(U;Y) + \left[I(V;X)-I(V;Y)\right]\\&= 1-H_2(\alpha\star \beta\star p) + \left[H_2(\alpha\star p)-H_2(\alpha)\right] \ .
\end{aligned}
\end{equation}

b) In this part, we let $V$ be degenerate and $\hat{X} = g(Y,V) = Y$. We then have $\mathbb{E}_0\left[d\left(X,\hat{X}\right)\right] = p$. Since in this case $I(V;X)-I(V;Y)=0$, we have 
\begin{equation}
\begin{aligned}
R_b &= I(U;Y) = 1-H_2(\alpha\star \beta\star p) \ .
\end{aligned}
\end{equation}

Now let $0 \leq D \leq p$ be given and say that $\theta, \alpha$ are such that $D = \theta\alpha + (1-\theta)p$. Since $R(D)$ is convex (for a given error exponent $E$),
\begin{equation}
\begin{aligned}
& R(E,D) = R(\theta\alpha + (1-\theta)p)
\\&\leq \theta R(\alpha) + (1-\theta)R(p)\\
&=\theta R_a + (1-\theta)R_b
\\&\leq 1-H_2(\alpha\star \beta\star p) + \theta\left[H_2(\alpha\star p)-H_2(\alpha)\right] \ .
\end{aligned}
\end{equation}
Thus, any triplet $(R,E,D)$ that complies with Theorem~\ref{Prop:BSC} is achievable through this scheme, and the proof of achievability is complete.

\subsection*{Converse proof}
Theorem~\ref{Prop:Optimality}, along with the development in \eqref{Eq:NewRate}, implies that the optimal region, for any specific example of hypothesis testing against independence, is comprised of two RVs, such that the Markov chain $U \mkv V \mkv X \mkv Y$ is respected. Moreover, it implies that with these optimal auxiliary RVs, the required rate is comprised of two independent parts -- one part dedicated to detection and the other to estimation. Thus, the proof of the converse to Theorem~\ref{Prop:BSC} can be divided, much like the proof of achievability, into two separate parts - one defining the trade-off between the rate and the error exponent, while the other defines the trade-off between the rate and the distortion.

Starting with the relation between the rate and the error exponent, Theorem~\ref{Prop:Optimality} implies that
\begin{equation}
\begin{aligned}
E &\leq I(U;Y) = H(Y)-H(Y|U) = 1-A \ ,
\end{aligned}
\end{equation}
while
\begin{equation}
R \geq 1-A+\theta\left[I(V;X)-I(V;Y)\right] \ ,
\end{equation}
with $A$ defined as $A \triangleq H(Y|U)$. Ignoring the second term in the expression for the rate, the trade-off between rate and error exponent is clear, and is given through $A$. Obviously, $A \leq H(Y) = 1$. In addition,
\begin{equation}
A \geq H_2\left(H_2^{-1}\left(H(X|U)\right)\star p\right) \ ,
\end{equation} 
which stems from Ms. Gerber's Lemma (see e.g. \cite{ElGamal-Kim-2011}). In order to allow the exploration of the entire region defined by the bounds over $A$, we define $\gamma \triangleq H_2^{-1}\left(H(X|U)\right)$. Thus, the trade-off between rate and error exponent becomes
\begin{equation}
\begin{aligned}
E &\leq 1-H_2(\gamma\star p) \ ,\\
R &\geq 1-H_2(\gamma\star p) +\theta\left[I(V;X)-I(V;Y)\right] \ .
\end{aligned}
\end{equation}

In the second part of the proof, it needs to be demonstrated that, once the decision $H_0$ has been (correctly) made, the optimal estimation region, defined by the rate-distortion relation $\min_{\mathbb{E}\left[d(X,\hat{X})\right] \leq D}\left[I(V;X)-I(Y;X)\right]$, is in agreement with Theorem~\ref{Prop:BSC}. This proof has already been given in \cite{Wyner-Ziv-1976} and is thus omitted from this work. Defining $V$ as the output of a BSC with cross-over probability $\alpha$ when $X$ is in the input of the channel, as was shown to be optimal in \cite{Wyner-Ziv-1976}, and keeping in mind the Markov chain implied by Theorem~\ref{Prop:Optimality}, it is clear that $\gamma = H^{-1}\left(H(X|U)\right) \geq \alpha$. Thus, $\gamma$ can be expressed as $\gamma = \alpha\star \beta$ for some $0 \leq \beta \leq \frac{1}{2}$, which completes the proof.

\section{Proof of Proposition~\ref{Prop:General}}\label{Appen:GeneralAchievability}
We now prove the achievability of the region offered in Proposition~\ref{Prop:General} for the joint detection and lossy compression problem, with general hypotheses. We start by describing the codebook, as well as encoding and decoding strategies, and followed by an analysis of error events under the proposed strategy.

\subsection*{Encoding and decoding strategy}
\emph{Codebook Construction:} For a given block-length $n$ we operate on a type-by-type basis. For each type $Q_X\in \mathcal{P}_n(\mathcal{X})$, fix a conditional type $Q_{U|X}^\star(Q_X)\in \mathcal{P}_n(\mathcal{U})$. %\in \mathcal{C}^n (Q_x,\mathcal{Y})$. 
Randomly and uniformly choose a set of codewords denoted by $\mathcal{C}^n_U(Q_X)$, from the resulting marginal type class $\cT_{Q_U^\star}^n(Q_X)$ which is induced by $Q_X$ and $Q_{U|X}^\star(Q_X)$. The size of $\mathcal{C}^n_U(Q_X)$ is an integer satisfying:
\begin{equation}\label{eq:BSize}
\begin{aligned}
\exp\big[{nI\big(Q_X;Q_{U|X}^\star(Q_X)\big)}\big]+&(|\mathcal{U}||\mathcal{X}|+2)\log(n+1)\\
\leq |\mathcal{C}_U^n&(Q_X)| \leq\\
\exp\big[{nI\big(Q_X;Q_{U|X}^\star(X)\big)}\big]+&(|\mathcal{U}||\mathcal{X}|+4)\log(n+1) \ ,
\end{aligned}
\end{equation}
where $\mathcal{C}_U^n(Q_X)$ is the codebook of the common message for source type $Q_X$.  Define $f_U: \cT_{Q_X}^n \to \mathcal{C}_U^n(Q_X)$, i.e., a function $f_U(\vct{x}^n)$ that determines the codeword sent by the encoder (node A) to the decoder (node B), as subsequently explained. We define $\vct{U}^n \triangleq f_U(\vct{X}^n)$. In addition, assign an index: $k(Q_{{X}}):\mathcal{P}_n(\mathcal{X}) \to\{1,\ldots,(n+1)^{|\mathcal{X}|}\}$ to each of the possible types of vectors $\vct{x}^n\in \mathcal{X}^n$.

In addition, let $V_0$ and $V_1$ be two RVs, designed to transmit a private message to the decoder. After making a decision about the common distribution controlling $X$ and $Y$, the decoder would use the appropriate private message in order to reconstruct the original sequence $\vct{x}$ (with distortion). As was the case when testing against independence as seen in Appendix~\ref{Apen:IndependenceAchievability},  the common distribution $Q_{UV|X} = Q_{U|X}Q_{V|UX}$ is chosen such that the Markov chains $U \-- V_0 \-- X \-- Y$ and $\bU \-- V_1 \-- \bX \-- \bY$ are respected. 

For each codeword $\vct{u}^n\in\mathcal{C}_U^n$, randomly generate $\exp{[nS_0]}$ sequences $\vct{v}_0^n(s_0)$, indexed with $s_0=[1:\exp{(nS_0)}]$, and $\exp{[nS_1]}$ sequences $\vct{v}_1^n(s_1)$, indexed with $s_1=[1:\exp{(nS_1)}]$, from the conditional typical sets $\cT_{[V_0|U]\delta}^n(\vct{u}^n)$ and  $\cT_{[V_1|\bU]\delta}^n(\vct{u}^n)$, respectively. Divide them into $\exp{(nR_0)}$ (respectively $\exp{(nR_1)}$) bins, such that each bin contains roughly $\exp{[n(S_0-R_0)]}$ (respectively $\exp{[n(S_1-R_1)]}$) sequences. In the remainder of this proof we only treat source reconstruction in case hypothesis $H_0$ was chosen, as the complementary case is completely symmetric.

\emph{Encoding:} Given a sequence $\vct{x}^n \in \cT_{Q_{X}}^n$, search for a sequence $\vct{u}^n\in\mathcal{C}^n_U(Q_{\vct{x}^n})$, i.e., in the codebook that belongs to the type $Q_{\vct{x}^n}$, such that $(\vct{u}^n,\vct{x}^n) \in \cT_{[UX]\delta}^n$. As a second step, look for a codeword $\vct{v}_0^n(s_0)$ such that $(\vct{v}_0^n(s_0),\vct{x}^n) \in \cT_{[V_0X|U]\delta}^n(\vct{u}^n)$ with the typicality measured according to the distribution induced by hypothesis $H_0$. Let $B_0(\vct{v}_0^n(\vct{x}^n,\vct{u}^n))$ denote the element (or ``bin'') to which $\vct{v}_0^n$ is mapped. Perform the same steps for the case where $H_1$ is the chosen hypothesis. 

The encoder's message then consists of four parts:
\begin{equation}
\begin{aligned}
\mathcal{M}_1 &= \{1,2,\dots,M_1 \triangleq \exp{(nR^\prime)}\} \ ,\\
\mathcal{M}_2 &= \big\{1,2,\dots,M_2 \triangleq (n+1)^{|\mathcal{X}|}\big\} \ ,\\
\mathcal{M}_3 &= \{1,2,\dots,M_3 \triangleq \exp{(nR_0)}\} \ ,\\
\mathcal{M}_4 &= \{1,2,\dots,M_3 \triangleq \exp{(nR_1)}\} \ ,\\
\mathcal{M} &= \mathcal{M}_1 \times \mathcal{M}_2 \times \mathcal{M}_3 \times \mathcal{M}_4 \ .
\end{aligned}
\end{equation}
The encoder sends the type of $\vct{x}^n$ which requires $|\mathcal{M}_2|$ values but with zero rate, and also $F(f_U(\vct{x}^n))$, as well as the respective bins for both private messages, $B_0(\vct{v}_0^n(\vct{x}^n,\vct{u}^n))$ and $B_1(\vct{v}_1^n(\vct{x}^n,\vct{u}^n))$, to be defined subsequently. There are two cases to consider:
\begin{itemize}
	\item[1] $\log|\mathcal{C}_U^n(Q_{\vct{x}^n})|<nR^\prime$, in which case we can map each member of ${C}_U^n(Q_{\vct{x}^n})$ to an element of $\mathcal{M}_1$ in a one-to-one manner.
	\item[2] $\log|\mathcal{C}_U^n(Q_{\vct{x}^n})|\geq nR^\prime$, in which case we assign each distinct member of ${C}_U^n(Q_{\vct{x}^n})$ to $\mathcal{M}_1$ uniformly at random.
\end{itemize}
Let $F(f_U(\vct{x}^n))$ denote the element to which $f_U(\vct{x}^n)$ is mapped. 
The encoder can be expressed mathematically as
\begin{equation}
\begin{aligned}
\Psi({x}) = &\big(F(f_U(\vct{x}^n)),k(Q_{\vct{x}^n})\\&\quad,B_0(\vct{v}_0^n(\vct{x}^n,\vct{u}^n)),B_1(\vct{v}^n_1(\vct{x}^n,\vct{u}^n))\big) \ ,
\end{aligned}
\end{equation}
for each $\vct{x}^n \in \cT_{Q_{\vct{x}^n}}^n$. 

\emph{Decoding:} The decoder first attempts to discover the word $\vct{u}^n$, by using the information sent from the encoder and  the observation vector $\vct{y}^n$:
\begin{itemize}
	\item If $\log|\mathcal{C}_U^n(Q_\vct{x})|<nR^\prime$ the codeword can be decoded without error;
	\item Otherwise $\log|\mathcal{C}_U^n(Q_\vct{x})|\geq nR^\prime$ the decoder receives a bin index and uses side information $\vct{y}^n$ to pick the best $\vct{u}^n$ in the bin. Given the bin number, the type $Q_{\vct{x}^n}$ and the side information $\vct{y}^n$, the decoder uses a minimal empirical entropy decoding\footnote{Note that since our chosen test is over empirical entropies, it does not matter at this stage which hypothesis is the true one, for the sake of choosing the sequence from the bin. After having retrieved a single sequence from the bin, the decoder can continue to perform HT by discarding the rest of the sequences in the bin and only using the chosen sequence.}, that is: 
	\begin{equation}
	\phi (F(f_U(\vct{x}^n)),Q_{\vct{x}^n},\vct{y}^n) = {\vct{\hat{u}}^n} \ ,
	\end{equation}
	if $H({\vct{\tilde{u}}^n}|\vct{y}^n) > H({\vct{\hat{u}}^n}|\vct{y}^n) $ for $\vct{\hat{u}}^n \in F(f_U(\vct{x}^n))$ and all $\vct{\tilde{u}}^n \in F(f_U(\vct{x}^n))$ with $\vct{\tilde{u}}^n \neq \vct{\hat{u}}^n$,  where
	$$
	H(\vct{\hat{u}}^n|\vct{y}^n) \triangleq  -\sum Q_{\vct{\hat{u}^n}\vct{y}^n}(a,b) \log Q_{\vct{\hat{u}^n}|\vct{y}^n}(a|b) 
	$$
	is the empirical entropy of the vector $\vct{\hat{u}^n}$ given the vector $\vct{y}^n$, and the sum is taken over all the letters in the alphabets of $U$ and $Y$.
\end{itemize}

As a second step, the decoder uses the private message --either $\vct{v}^n_0$ or $\vct{v}^n_1$-- destined for the case of the current hypothesis  in order to estimate $\vct{x}^n$, with distortion $D_0$ or $D_1$, respectively. Assume hypothesis $H_0$ is in effect,  it searches for a single sequence $\vct{\hat{v}}^n_0 \in B_0(\vct{v}_0^n(\vct{x}^n ,\vct{u}^n))$ such that $\vct{\hat{v}}^n_0(s_0) \in \cT_{[V_0|UY]\delta}(\vct{u}^n\vct{y}^n)$. If it finds no such sequence it declares an error during the reconstruction. If it finds more than one, it chooses one sequence at random.

\subsection*{Error probability of the testing step}
We now show that, for the detection part, the exponential rate of decay of the error of the second type, under a fixed constraint over the error of the first type, is not smaller than the value claimed by Proposition~\ref{Prop:General}. The analysis of possible errors at the encoder's side stays identical to the one done in the proof of Theorem~\ref{Prop:Optimality} in Appendix~\ref{Apen:IndependenceAchievability} (note that we assume the $P_X(x) = P_{\bX}(x)$, without which the analysis of the encoder's side, with an emphasis on the codebook construction, might become more involved). Note also that when a problem does arise during encoding, our proposed scheme calls for an error message which prompts node $B$ to declare $H_1$. Thus, the influence of such errors is only on the error probability of Type I, and not on the error exponent of Type II. We concentrate in this analysis on possible errors at the decoder's side. Define two error events: First, let
\begin{equation}
\cB_6 \triangleq \{\vct{u}^n \neq F(f_U(\vct{x}^n)) \}
\end{equation}
be the event that the chosen sequence from the bin at the decoder is different from the original sequence sent by the encoder. Then, define $\cB_7$ to be the event of erroneous detection despite using the correct sequence. We denote the probabilities of events $\cB_6$ and $\cB_7$ by $P_r^{(n)}$ and $P_d^{(n)}$, respectively.
Using the union bound, the probability of error in detection can be bounded by
\begin{equation}
P^{(n)}_e \leq P^{(n)}_{r} + P^{(n)}_{d} \ .
\end{equation}
%Throughout this analysis, we ignore the possibility of failure at the transmission's side, as was done in \cite{Shimokawa-han-amari-1994}. The reason we can do so is that when such a failure occurs (which does happen rarely, thanks to the AEP), the transmitter can always signal the receiver to choose $H_1$. This strategy will prevent any influence on the error exponent of the second type, while not hurting the constraint over the probability of error of the first type, when $n$ is large enough.

\emph{Evaluation of $P_r^{(n)}:$} 
We evaluate the probability that node $B$ chooses the wrong sequence from the bin under the suggested encoding and decoding schemes. Our evaluation is reliant on the method of types \cite{Csiszar-1998}, and is specifically inspired by the techniques used in \cite[Appendix C]{Kelly-Wagner-2012}. We first evaluate $P^{(n)}_r$ for a finite block-length $n$ and then use a continuity argument to show that in the limit of $n \to \infty$,
\begin{equation}
\begin{aligned}
&-\frac{1}{n}\log P^{(n)}_r \leq G(Q_{UXY},Q_{X},Q_{Y},R^\prime) %= \\
%&\quad \begin{cases}
%\min\limits_{i = \{0,1\}} \mathcal{D}\big(Q_{UXY}\|P_{UXY_i}\big) + \left[R^\prime-I\big(X;U\big) + I\big(Y;U\big)\right]^+ & I\big(Q_X;Q_{U|X}\big) > R^\prime \\
%\infty & \text{else.}
%\end{cases}
\end{aligned} \ ,
\end{equation} 
where the function $G$ is the one given in \eqref{eq:G}.

Since choosing the wrong sequence can only happen in case binning is used, we are only interested in the following subset of the set of all possible sequences:
\begin{equation}
\begin{aligned}
\mathcal{A}_n &= \Big\{(\vct{u}^n,\vct{x}^n,\vct{y}^n)\in \mathcal{U}^n\times\mathcal{X}^n\times \mathcal{Y}^n  \\& \quad\big| \, \vct{u}^n \in T_{Q_{{U}|{X}}^\star}^n(Q_{\vct{x}^n}) \,,\, \log|\mathcal{C}_U^n(Q_{\vct{x}^n})| \geq nR\Big\} \ .
\end{aligned}
\end{equation}
We first evaluate the probability of choosing the wrong sequence within the set $\mathcal{A}_n$ by using the following lemma.

\begin{lemma}\label{LemmaBinning} Let $(\vct{u}^n,\vct{x}^n,\vct{y}^n) \in \mathcal{A}_n$ and let  $\mathcal{B}_8$ be the event that $\vct{u}^n \neq \phi(\psi(\vct{x}^n),\vct{y}^n)$. 
Provided that $\log|\mathcal{C}_U^n(Q_{\vct{x}^n})|\geq nR$, then
	\begin{equation}\label{Eq:LemmaProbability}
	\begin{aligned}
	\Pr &\left(\mathcal{B}_8|\vct{U}^n=\vct{u}^n,\vct{X}^n = \vct{x}^n, \vct{Y}^n=\vct{y}^n\right) \\&\leq \exp\big[ -n\left(R-J(Q_{\vct{u}^n\vct{x}^n\vct{y}^n})-\delta_n\right) \big] \ ,
	\end{aligned}
	\end{equation} 
	with
	\begin{equation}
	\begin{aligned}
		& J\big(Q_{\vct{u}^n\vct{x}^n\vct{y}^n}\big) \\&\triangleq I\big(Q_{{\vct{x}^n}};Q_{U|X}^\star(Q_{{\vct{x}^n}})\big) - I\big(Q_{{\vct{u}^n}|{\vct{y}^n}};Q_{{\vct{y}^n}}\big)
	\end{aligned}
	\end{equation}
	and
	\begin{equation}
	\delta_n \triangleq \frac{1}{n} \log(n+1)^{|\mathcal{U}|(1+|\mathcal{X}|+|\mathcal{Y}|) +4} \ .
	\end{equation}
	The probability in \eqref{Eq:LemmaProbability} is taken over the choice of the codebook in use.
\end{lemma}

Before proving Lemma~\ref{LemmaBinning}, we recall the following result 
from~\cite[Lemma 12]{Kelly-Wagner-2012}. 
\begin{lemma}\label{lemma-Kelly-Wagner}
	For all strings $(\vct{u},\vct{x})$ such that $\vct{u} \in T_{Q_{U}^{\star}}^n$,
	\begin{equation}
	\begin{aligned}
		\Pr  &(\vct{u} \in \mathcal{C}_U^n(Q_{\vct{x}^n}))\\& \leq (n+1)^{\|\mathcal{U}\|(1+\|\mathcal{X}\|)+4}\\&\times\exp \left[n\left(I(Q_{\vct{x}^n};Q_{U|X}^\star(Q_{\vct{x}^n}))-H(Q_{{\vct{u}^n}})\right)\right].
	\end{aligned}
	\end{equation}
	\end{lemma}

\begin{IEEEproof}[Proof (Lemma~\ref{LemmaBinning})]
	Let $\mathcal{S}(\vct{u}^n|\vct{y}^n)$ be the set that includes all sequences ${\vct{\tilde{u}}^n}$, such that ${\vct{\tilde{u}}^n}$ has the same type as $\vct{u}$ and $H(\vct{\tilde{u}}^n|\vct{y}^n) \leq H(\vct{u}^n|\vct{y}^n)$. Then
	\begin{subequations}
		\begin{align}
		&\Pr \left(\mathcal{B}_8|\vct{U}^n=\vct{u}^n,\vct{X}^n = \vct{x}^n, \vct{Y}^n=\vct{y}^n\right)\nonumber \\ 
		&\leq \sum %\limits_{\substack{{\vct{\tilde{u}}^n} \in \mathcal{S}(\vct{u}^n|\vct{y}^n)\\ {\vct{\tilde{u}}^n} \neq \vct{u}^n}} 
		{\Pr} \big( {\vct{\tilde{u}}^n} \in \mathcal{C}_U^n(Q_{\vct{x}^n}), \{F(\vct{\tilde{u}}^n)=F(\vct{u}^n)\} | \\&\qquad\qquad\vct{U}^n=\vct{u}^n,\vct{X}^n = \vct{x}, \vct{Y}^n=\vct{y}\big)\nonumber\\
			&\leq \sum %\limits_{\substack{{\vct{\tilde{u}}^n} \in \mathcal{S}(\vct{u}^n|\vct{y}^n)\\ {\vct{\tilde{u}}^n} \neq \vct{u}^n}} 
		{\Pr} \big(\vct{\tilde{u}}^n \in \mathcal{C}_U^n(Q_{\vct{x}^n}) | \vct{X}^n = \vct{x}^n, \vct{Y}^n=\vct{y}^n\big)\nonumber\\&\qquad\qquad\times {\Pr} \big(\{F(\vct{\tilde{u}}^n)=F(\vct{u}^n)\}\big) \label{Eq:B81}\\
		&\leq \sum %\limits_{\substack{{\vct{\tilde{u}}^n} \in \mathcal{S}(\vct{u}^n|\vct{y}^n)\\ {\vct{\tilde{u}}^n} \neq \vct{u}^n}}  
		(n+1)^{|\mathcal{U}|(1+|\mathcal{X}|)+4}\nonumber\\&\times\exp\big[{n\left(I(Q_{\vct{x}^n};Q_{U|X}^\star(Q_{\vct{x}^n}))-H(Q_{{\vct{u}^n}})\right)}\big]\frac{1}{M_1}\label{Eq:B82}
			\end{align}
	\end{subequations}
	
	\begin{subequations}
		\begin{align}			
		&\leq (n+1)^{|\mathcal{U}||\mathcal{Y}|}\exp\big[{nH(Q_{{\vct{u}^n}|{\vct{y}^n}}|Q_{{\vct{y}^n}})}\big]\frac{1}{M_1}\nonumber\\&\times(n+1)^{|\mathcal{U}|(1+|\mathcal{X}|)+4} \label{Eq:B83}\\ &\times\exp\big[{n\left(I(Q_{\vct{x}^n};Q_{U|X}^\star(Q_{\vct{x}^n}))-H(Q_{\vct{u}^n})\right)}\big]\nonumber\\
		&= (n+1)^{|\mathcal{U}|(1+|\mathcal{X}|+|\mathcal{Y}|)+4}\nonumber\\&\times\exp\left[-n\left(R - H(Q_{{\vct{u}^n}|{\vct{y}^n}}|Q_{{\vct{y}^n}}) +H(Q_{{\vct{u}^n}}) \right.\right.\nonumber\\&\left.\left.\qquad\quad- I(Q_{\vct{x}^n};Q_{U|X}^\star(Q_{\vct{x}^n}))\right)\right]\label{Eq:B84}\\
		&= (n+1)^{|\mathcal{U}|(1+|\mathcal{X}|+|\mathcal{Y}|)+4} \nonumber\\&\times\exp\left[-n\left(R + I(Q_{{\vct{u}^n}|{\vct{y}^n}};Q_{{\vct{y}^n}})) \right.\right.\nonumber\\&\left.\left.\qquad\quad- I(Q_{\vct{x}^n};Q_{U|X}^\star(Q_{\vct{x}^n}))\right)\right]\\
		&\triangleq (n+1)^{|\mathcal{U}|(1+|\mathcal{X}|+|\mathcal{Y}|)+4} \nonumber\\&\times\exp\big[-n\left(R - J(Q_{\vct{u}^n\vct{x}^n\vct{y}^n})\right)\big]\\
		&\leq \exp\big[-n\left(R - J(Q_{\vct{u}^n\vct{x}^n\vct{y}^n}) - \delta_n\right)\big] \ , \label{Eq:B85}
		\end{align}
	\end{subequations}
	with $\delta_n$ as defined above, and the sums are all taken over the set $\vct{\tilde{u}}^n \in \mathcal{S}(\vct{u}^n|\vct{y}^n)$, $\vct{\tilde{u}}^n \neq \vct{u}^n$. Here, the probability ${\Pr} \left(\vct {\tilde{u}}^n \in \mathcal{C}_U^n(Q_{x^n})\right)$ is over the choice of the codebook. Inequality \eqref{Eq:B81} stems from the codebook construction, which divides sequences into bins randomly and independently.  Inequality \eqref{Eq:B82} is due to Lemma~\ref{lemma-Kelly-Wagner}, which applies here with slight notation changes (see at the end of this proof), and to the upper bound over the size of $\mathcal{C}_U^n(Q_{\vct{x}^n})$, given in \eqref{eq:BSize}. Inequality \eqref{Eq:B83} is due to Lemma~\ref{Lemma:EmpiricalEntropy}. Finally, equality \eqref{Eq:B84} is due to the definition of $M_1$ and \eqref{Eq:B85} stems from the fact that $\Pr \left(\mathcal{B}_8|\vct{U}^n=\vct{u}^n,\vct{X}^n = \vct{x}^n, \vct{Y}^n=\vct{y}^n\right) \leq 1$ and the definition of $\delta_n$.

\end{IEEEproof}

We now bound the probability of error in choosing the right sequence in the bin $P_r^{(n)}$, for a finite block-length $n$:
\begin{subequations}
	\begin{align}
	& P_r^{(n)} = \Pr\left(\{\vct{u}^n \neq F(f_U(\vct{x}^n)) \}\right)\\
	&\leq \sum %\limits_{(\vct{u}^n,\vct{x}^n,\vct{y}^n) \in \mathcal{A}_n} 
	\text{Pr} \left(\mathcal{B}_8|\vct{U}^n=\vct{u},\vct{X}^n = \vct{x}, \vct{Y}^n=\vct{y}\right) \nonumber\\&\quad\times\text{Pr} \left(\vct{U} = \vct{u},\vct {X} = \vct{x}, \vct{Y} = \vct{y}\right)\\
	&\leq \sum %\limits_{(\vct{u}^n,\vct{x}^n,\vct{y}^n) \in \mathcal{A}_n} 
	\exp\big[{-n\left(R - J(Q_{\vct{u}^n\vct{x}^n\vct{y}^n}) - \delta_n\right)}\big] \nonumber\\&\quad\times P_{XY}^n(\vct{x}^n,\vct{y}^n) \frac{1}{|\cT_{Q_{U|X}^\star}^n(Q_{\vct{x}^n})|} \ .\label{Eq:UniGeneralPr}
	\end{align}
\end{subequations}
Here, claim \eqref{Eq:UniGeneralPr} is derived from Lemma~\ref{LemmaBinning}. Note the slight abuse of notation here, where $P_{XY}^n(\vct{x}^n,\vct{y}^n)$ in \eqref{Eq:UniGeneralPr} refers to the \emph{real distribution} controlling the RVs, and can thus actually be, according to the true hypothesis, wither $P_{XY}^n(\vct{x}^n,\vct{y}^n)$ or $P_{\bX\bY}^n(\vct{x}^n,\vct{y}^n)$. The probability of choosing a specific sequence $\vct{u}^n$ given both source sequences $\vct{x}^n$ and $\vct{y}^n$ stems from averaging over the code. We can now change the expression to sum first on types and then on sequences within each type class. In order to transform our summation over a set of sequences $\mathcal{A}_n$ into a summation over a set of types (and only then over the sequences within each type) we define the following set of types:
\begin{equation}
\begin{aligned}
&\mathcal{D}(Q_X,Q_Y) \\&=\big \{Q_{UXY}\in\mathcal{P}_n(\mathcal{U}\times\mathcal{X}\times\mathcal{Y}): Q_{U|X} = Q_{U|X}^\star(Q_X), \\& \qquad\qquad\qquad\qquad\qquad\qquad \log|\mathcal{C}_U^n(Q_X)| \geq nR\big\} \ .
\end{aligned}
\end{equation}
The probability of error in selecting the sequence can thus be bound by \eqref{eq:BoundPr}, at the top of the next page.
	\begin{table*}
\begin{equation}\normalsize
\begin{aligned}\label{eq:BoundPr}
P_r^{(n)} &\leq \sum\limits_{Q_X,Q_Y} \left[\sum\limits_{Q_{UXY}\in \mathcal{D}(Q_X,Q_Y)} \,\,\sum\limits_{(\vct{u}^n,\vct{x}^n,\vct{y}^n)\in \cT_{Q_{UXY}}^n} \frac{P_{XY}^n(\vct{x}^n,\vct{y}^n)}{|\cT_{Q_{U|X}^\star}^n( Q_{\vct{x}^n})|} \exp\big[{-n\big(R - J(Q_{\vct{u}^n\vct{x}^n\vct{y}^n}) - \delta_n\big)}\big]\right] \ .
\end{aligned}
\end{equation}
		%\protect\caption{Long equation}
	\end{table*}
%\begin{equation}
%\begin{aligned}
%P_r^{(n)} &\leq \sum\limits_{Q_X,Q_Y} \left[\sum\limits_{Q_{UXY}\in \mathcal{D}(Q_X,Q_Y)} \,\,\sum\limits_{(\vct{u}^n,\vct{x}^n,\vct{y}^n)\in \cT_{Q_{UXY}}^n} \frac{P_{XY}^n(\vct{x}^n,\vct{y}^n)}{|\cT_{Q_{U|X}^\star}^n( Q_{\vct{x}^n})|} \exp\big[{-n\big(R - J(Q_{\vct{u}^n\vct{x}^n\vct{y}^n}) - \delta_n\big)}\big]\right] \ .
%\end{aligned}
%\end{equation}

In the case of distributed HT, the probability of the source sequences $(\vct{x}^n,\vct{y}^n)$ is unknown, since the sequences can be created by one of two possible distributions. We thus bound the probability of the observed sources by 
\begin{equation}
\begin{aligned}
& P_{XY}^n(\vct{x}^n,\vct{y}^n) \leq \max\{P_{XY}(\vct{x}^n,\vct{y}^n),P_{\bX\bY}(\vct{x}^n,\vct{y}^n) \}\\
&= \max\limits_{i= \{0,1\}}\big\{\exp\big[{-n\left(\mathcal{D}(Q_{XY}\| P_{XY_i}) + H(Q_{XY})\right)}\big] \big\}\\
%&=& \max\limits_{i \in \{0,1\}} \big\{\exp\big[-n\left(\mathcal{D}(Q_{XY}\| P_{XY_i})\right)\big]\big\} \exp\big[{-nH(Q_{XY})}\big]\\
&= \exp\left[{-n\left(\min\limits_{i = \{0,1\}} \mathcal{D}(Q_{XY}\| P_{XY_i}) + H(Q_{XY})\right)} \right]\ ,
\end{aligned}
\end{equation}
where, in accordance to the notation of Proposition~\ref{Prop:General}, we use the subscript $i$ in order to differentiate between $P_{XY}$ and $P_{\bX\bY}$. 
Using the following facts detailed in Lemma~\ref{lemma:SizeTypicalSet},
\begin{subequations}
	\begin{align}
	|\cT_{Q_{UXY}}^n| &\leq \exp\big[{n(H(Q_{UXY}))}\big] \nonumber\\&\leq \exp\big({n\log |\mathcal{U}||\mathcal{X}||\mathcal{Y}|} \big)\ ,\\
	|\cT_{Q_{U|X}}^n| &\geq (n+1)^{-|\mathcal{U}||\mathcal{X}|}\exp\big[{n\left(H(Q_{U|X}|Q_X)\right)}\big] \ ,
	\end{align}
\end{subequations}
we obtain from \eqref{eq:BoundPr} that
\begin{equation}
\begin{aligned}
%P_r^{(n)} &\leq \sum\limits_{Q_X\in \mathcal{P}_n(\mathcal{X})} \sum\limits_{Q_Y\in \mathcal{P}_n(\mathcal{Y})} \left[\sum\limits_{Q_{UXY}\in \mathcal{D}(Q_X,Q_Y)} \exp\left[{-n\left(\min\limits_{i= \{0,1\}} \mathcal{D}(Q_{XY}\| P_{XY_i}) + H(Q_{XY})\right)} \right]\times\right.\nonumber\\
%& \quad(n+1)^{|\mathcal{U}||\mathcal{X}|}\exp\big[{n H(Q_{U|X}|Q_X) }\big]\times\exp\big[{n H(Q_{UXY})}\big] \exp\big[{-n\left(R - J(Q_{UXY}) - \delta_n\right)}\big]\nonumber\\
&\leq \sum\limits_{Q_X\in \mathcal{P}_n(\mathcal{X})} \sum\limits_{Q_Y\in \mathcal{P}_n(\mathcal{Y})}\sum\limits_{Q_{UXY}\in \mathcal{D}(Q_X,Q_Y)} \\&\qquad\exp\big[{-n\left( \Gamma + R - J(Q_{UXY}) - \delta_n \right)}\big] \ ,
\end{aligned}
\end{equation}
with $\Gamma$ satisfying:
\begin{equation}
\begin{aligned}
\Gamma &= \min\limits_{i =\{0,1\}} \mathcal{D}(Q_{XY}\|P_{XY_i}) + H(Q_{XY})\\& + H(Q_{U|X}|Q_X) - H(Q_{UXY})\\
&= \min\limits_{i = \{0,1\}}\mathcal{D}(Q_{XY}\|P_{XY_i})+ H(Q_{U|X}|Q_X) \\&- H(Q_{U|XY}|Q_{XY})\\
&= \min\limits_{i = \{0,1\}}\sum\limits_{\substack{x \in \mathcal{X}\\ y \in \mathcal{Y}}} Q_{XY}(x,y)\log\frac{Q_{XY}(x,y)}{P_{XY_i}(x,y)} \\&- \sum\limits_{\substack{u \in \mathcal{U}\\ x \in \mathcal{X}}} Q_{UX}(u,x)\log\frac{Q_{UX}(u,x)}{Q_{X}(x)}\\
& + 
\sum\limits_{\substack{u \in \mathcal{U}\\x \in \mathcal{X}\\ y \in \mathcal{Y}}}  Q_{UXY}(u,x,y)\log\frac{Q_{UXY}(u,x,y)}{Q_{XY}(x,y)}\\
%&= \min\limits_{i \in \{0,1\}}\Biggl\{\sum\limits_{\substack{u \in \mathcal{U}\\x \in \mathcal{X}\\ y \in \mathcal{Y}}} Q_{UXY}(u,x,y)\log\frac{Q_{XY}(x,y)}{P_{XY_i}(x,y)}\frac{Q_{X}(x)}{Q_{UX}(u,x)}\frac{Q_{UXY}(u,x,y)}{Q_{XY}(x,y)}\Biggr\}\\
%&= \min\limits_{i = \{0,1\}}\Biggl\{\sum\limits_{\substack{u \in \mathcal{U}\\x \in \mathcal{X}\\ y \in \mathcal{Y}}} Q_{UXY}(u,x,y)\log \frac{Q_{UXY}(u,x,y)}{P_{XY_i}(x,y)Q_{U|X}(u|x)}\Biggr\}\\
&= \min\limits_{i= \{0,1\}} \mathcal{D}(Q_{UXY}\|P_{XY_i}Q_{U|X}) \ .
\end{aligned}
\end{equation}
The probability of error in bin decoding can thus be concluded to satisfy
\begin{equation}
\begin{aligned}
P_r^{(n)}& \leq \sum\limits_{Q_X \in \mathcal{P}_n(\mathcal{X})} \sum\limits_{Q_Y \in \mathcal{P}_n(\mathcal{Y})}\sum\limits_{Q_{UXY}\in \mathcal{D}(Q_X,Q_Y)}\\ 
&\exp\left[-n\left( \min\limits_{i = \{0,1\}} \mathcal{D}(Q_{UXY}\|P_{XY_i}Q_{U|X}) \right.\right.\\&\left.\left.\qquad+ R - J(Q_{UXY}) - \delta_n \right)\right] \ .
\end{aligned}
\end{equation}
We may now upper bound the summations by maximizing over the types and optimizing over the choice of the of the test channel $Q_{U|X}^\star$. We optimize to then obtain:
\begin{equation}
\begin{aligned}
P_r^{(n)} & \leq |\mathcal{P}_n(\mathcal{X})|\max\limits_{Q_X}\min\limits_{Q_{U|X}^\star} |\mathcal{P}_n(\mathcal{Y})|\max\limits_{Q_Y} |\mathcal{P}_n(\mathcal{U}\times\mathcal{X}\times\mathcal{Y})| \\&\quad \!\!\!\max\limits_{\substack{Q_{UXY}\\Q_{U|X}=Q_{U|X}^\star}} \exp\Big\{{-nG_n\left[Q_{UXY},Q_X,      Q_Y,R\right]}\Big\} \ .
\end{aligned}
\end{equation}
Thus,
\begin{equation}
\begin{aligned}
\frac{1}{n}\log P_r^{(n)}  & \leq - \min\limits_{Q_X\in \mathcal{P}_n(\mathcal{X})}\max\limits_{Q_{U|X}^\star(Q_X)} \min\limits_{  Q_Y\in \mathcal{P}_n(\mathcal{Y})} \min\limits_{\substack{Q_{UXY}\\Q_{U|X}=Q_{U|X}^\star}}\\& \quad G_n\left[Q_{UXY},Q_X,  Q_Y,R \right] \nonumber \\ 
&\quad\times\log \left(|\mathcal{P}_n(\mathcal{X})||\mathcal{P}_n(\mathcal{Y})||\mathcal{P}_n(\mathcal{U}\times\mathcal{X}\times\mathcal{Y})|\right)
\end{aligned} 
\end{equation}
with the function $G_n\left[Q_{UXY},Q_X, Q_Y,R \right]$ defined in \eqref{eq:Gn} at the top of the next page.
	\begin{table*}
\begin{equation}\normalsize
\begin{aligned}
G_n\left[Q_{UXY},Q_X,Q_Y,R \right] &=
&\begin{cases}
\begin{aligned}
&\min\limits_{i = \{0,1\}} \mathcal{D}(Q_{UXY}\|P_{XY_i}Q_{U|X}) \\ &+ \left[R - I(Q_{X};Q_{U|X}^\star) + I(Q_{Y};Q_{U|Y}^\star)\right]
\end{aligned} & I(Q_{X};Q_{U|X}^\star) > R\\
+\infty & \text{else}\ .
\end{cases}
\end{aligned}\label{eq:Gn}
\end{equation}
		%\protect\caption{Long equation}
	\end{table*}
%\begin{equation}
%\begin{aligned}
%G_n\left[Q_{UXY},Q_X,Q_Y,R \right] &=
%&\begin{cases}
%\begin{aligned}
%&\min\limits_{i = \{0,1\}} \mathcal{D}(Q_{UXY}\|P_{XY_i}Q_{U|X}) \\ &+ \left[R - I(Q_{X};Q_{U|X}^\star) + I(Q_{Y};Q_{U|Y}^\star)\right]
%\end{aligned} & I(Q_{X};Q_{U|X}^\star) > R\\
%+\infty & \text{else}\ .
%\end{cases}
%\end{aligned}
%\end{equation}
The cardinalities can be absorbed inside the exponent and become insignificant as $n \to \infty$. From continuity arguments under discrete alphabets, it is made clear that \cite[Lemma 14]{Kelly-Wagner-2012}:
\begin{equation}
\begin{aligned}
& P_r^{(n)} \leq \inf\limits_{Q_X\in \mathcal{P}(\mathcal{X})}\sup\limits_{Q_{U|X}^\star(Q_X)\in\mathcal{P}(\mathcal{U})} \inf\limits_{ Q_Y\in \mathcal{P}(\mathcal{Y})} \inf\limits_{\substack{Q_{UXY}\in \mathcal{P}(\mathcal{U}\times\mathcal{X}\times\mathcal{Y})\\Q_{U|X}=Q_{U|X}^\star}}\\&\quad \quad \quad \quad \quad G\left[Q_{UXY},Q_X,  Q_Y,R\right] \ ,
\end{aligned}
\end{equation}
where all the optimization steps are now being taken over \emph{probability distributions}, and $G$ is as defined in Proposition~\ref{Prop:General}.

\emph{Evaluation of $P_d^{(n)}$:} We now study the Type II error probability of detection, under the assumption that the right sequence has been correctly extracted from the bin. The probability that, given the right sequence $\vct{u}^n$, node B makes a wrong decision was investigated in detail in~\cite{Han-1987}, using the method of types \cite{Csiszar-1998}, as well as properties of types and typical sequences, detailed in Appendix~\ref{Apen:typicality} of this paper. That result, however, is dependent on a specific codebook, conceived to allow detection with high probability. As we use a random codebook in our scheme, it is essential to adapt the method of~\cite{Han-1987}. We give here a general description of this adaptation. 

We propose here a slight modification to~\cite{Han-1987}. Intuitively, since we investigate the exponential decay of $\beta_n$ while only enforcing a fixed upper bound on $\alpha_n$, we show that the penalty of replacing the codebook construction in~\cite{Han-1987} with random coding can be fully absorbed into $\alpha_n$, leaving the error exponent result of $\beta_n$ unmodified. Nevertheless, $\alpha_n$ can still be shown to approach $0$ as $n$ grows, which indicates that any constraint $\alpha_n \leq \epsilon$ can be fulfilled, for $n$ large enough and $\epsilon > 0$. For the given codebook, define 
\begin{equation}
\begin{aligned}
\mathcal{L}(Q^\star_{UX},Q_{UY}^\star) = \Big\{& P_{\tilde{U}\tilde{X}\tilde{Y}} \in \mathcal{P}(\mathcal{U} \times \mathcal{X} \times \mathcal{Y}) \,: \\ & P_{\tilde{U}\tilde{X}}(u,x) = Q^\star_{UX}(u,x),\\ & P_{\tilde{U}\tilde{Y}}(u,y) = Q_{UY}^\star(u,y),  \forall \,(u,x,y)\Big\} \ ,
\end{aligned}
\end{equation} 
to be the set of all triplets of auxiliary RVs such that the marginal distribution of each pair ($U,X$) and ($U,Y$) is maintained. Similarly to \cite{Han-1987}, it is not difficult to show that, for the codebook described above,
\begin{equation}\label{Eq:Lower}
\theta_L(R) \triangleq \min \limits_{\tilde{U}\tilde{X}\tilde{Y} \in \mathcal{L}(Q^\star_{UX},Q_{UY}^\star)} \mathcal{D}(P_{\tilde{U}\tilde{X}\tilde{Y}}\|P_{\bU\bX\bY}) 
\end{equation}
provides a lower bound to the error probability of the second type, after the correct sequence has been recovered from the bin, and under a fixed error probability of the first type.

From the construction of the codebook (specifically the size of the set $\mathcal{C}_U^n(Q_{\vct{x}^n})$), it can be seen that the number of sequences in the codebook \emph{per type of ${X}$} complies with $M = \exp\big[{n(I(Q_{\vct{x}^n};Q^\star_{U|X}(Q_{\vct{x}^n}))+\eta)}\big]$. Given  a sequence $\vct{x}^n$, search for a sequence $\vct{u}_i$ in the codebook that belongs to the type of $\vct{x}^n$, such that $(\vct{u}_i^n,\vct{x}^n) \in \cT_{[UX]\delta}^n$ and send its index (or bin number, depending on the type of $\vct{x}^n$) to the receiver. As we only consider here the error event where the wrong hypothesis is chosen despite the correct sequence is used, we ignore errors in choosing the correct sequence from the bin, in case binning has occurred, for the sake of this analysis.  If there is more than one such sequence choose randomly. If there is no such sequence in the codebook, send an error message. At the decoder (node B), if $(\vct{u}^n_i,\vct{y}^n) \in \cT_{[UY]\delta}^n$ (notice that typicality here is checked only under hypothesis $H_0$) declare $H_0$. In any other case (including the case an error message was received) declare $H_1$. This choice allows us to ``push'' the penalty of not using the code proposed in \cite[Lemma 4]{Han-1987} into $\alpha_n$ (which, when $n \to \infty$ can still be bounded by any fixed $\epsilon >0$), thus leaving the evaluation of $\beta_n$ unchanged, as shown subsequently.

\emph{Evaluation of $\alpha_n$:} An error of the first type occurs if for $n$ i.i.d. samples $(\vct{x}^n,\vct{y}^n) \sim P_{XY}(x,y)$ (hypothesis $H_0$ holds) the decoder declares $H_1$. According to the proposed coding schemes, two possible events can induce the decoder to such an error. The first is given by 
\begin{equation}
\text{(i)} \quad \cB_9 \triangleq \{\nexists \text{ $i$ such that } (\vct{u}_i^n,\vct{x}^n) \in \cT_{[UX]\delta}^n\} \ .
\end{equation}
Assuming without loss of generality that the sequence $\vct{u}_1^n$ was chosen and sent from node A, the second relevant error event is:
\begin{equation}
\text{(ii)} \quad \cB_{10} \triangleq \{H_0\text{ is true and } (\vct{u}_1^n,\vct{y}^n) \notin \cT_{[UY]\delta}^n\} \ .
\end{equation}
From the union bound, it is obvious that:
\begin{equation}
\alpha_n \leq \Pr(\cB_9) + \Pr(\cB_{10}\cap \cB_9^c) \ .
\end{equation}
Through the AEP it is easy to conclude that both of these probabilities approach zero when $n \to \infty$. Thus, for $n$ large enough one can conclude that $\alpha_n \leq \epsilon$ for any fixed $\epsilon >0$. 

\emph{Evaluation of $\beta_n$:} The error of the second type can be defined by  a single event:
\begin{equation}
\quad \cB_{11} \triangleq \{H_1\text{ is true and } (\vct{u}_1^n,\vct{y}^n) \in \cT_{[UY]\delta}^n\} \ .
\end{equation}
The analysis of $\beta_n$ is identical to what was done in \cite{Han-1987}. One important difference, however, is that by defining
\begin{equation}
\mathscr{C}_i \triangleq \Big\{\vct{x}^n\in\mathcal{X}^n: \quad (\vct{u}_i^n,\vct{x}^n) \in \mathcal{T}_{[UX]\delta}^n\Big\} \ ,
\end{equation}
the sets $\mathscr{C}_i$ are not necessarily disjoint. This, however, does not change the calculations by following same steps as in~\cite{Han-1987}. 
 
 \subsection*{Source reconstruction}
As a final step, we demonstrate the achievability of the estimation part in Proposition~\ref{Prop:General}, for the case where hypothesis $H_0$ is chosen (the case of hypothesis $H_1$ is symmetric). 
\begin{Remark}
	Note that the achievable scheme used here in order to prove Proposition~\ref{Prop:General} ensures that $\alpha_n \to 0$ when $n \to \infty$, despite this not being a requirement. This is crucial in order for the following analysis, done for hypothesis $H_0$, to be applicable equivalently also for hypothesis $H_1$.
\end{Remark}
Denoting by $\cB_{12}$ the event ``an error occurred during encoding or decoding, under the correct decision $H_0$'', we expand its probability as follows: $\text{Pr}(\cB_{12}) \leq P^\prime+P^{\prime\prime}$, with $P^{\prime}$ being the probability that no codeword $\vct{v}_0^n(s_0)$ could be found in the codebook for the given sequence $\vct{x}^n$ and the chosen sequence $\vct{u}^n$, and $P^{\prime\prime}$ being the probability that a different codeword in the same bin is compatible with $\vct{y}^n$ and $\vct{u}^n$. 

Using standard arguments, both error probabilities can be bounded as follows:
\begin{equation}
\begin{aligned}
& P^\prime \triangleq \Pr\{\nexists \,s_0=[1:\exp{(nS_0)}] \, \\&\qquad\text{ s.t. } \,(\vct{V}_0^n(s_0),\vct{X}^n) \in \cT_{[V_0X|U]\delta}^n(\vct{u}^n)\}\\
&\leq   \Pr\{({V}_0^n,{X}^n) \notin \cT_{[V_0X|U]\delta}^n(\vct{u}^n)|\\&\qquad {V}^n \in \cT_{[V_0|U]\delta}^n(\vct{u}^n), {X}^n \in \cT_{[X|U]\delta}^n(\vct{u}^n) \}^{\exp{(nS_0)}}\\
&\leq \exp \left\{-\exp{[nS_0]}\right.\\& \left.\qquad\times\exp\big[-n( I(X;V_0|U)+\eta_n^{(1)}) \big] \right\}\\
&=\exp\big\{-\exp{\big[-n\big(I(X;V_0|U) - S_0 + \eta_n^{(1)}\big)\big]}\big\}\ .
\end{aligned}
\end{equation}
Thus, $P^{\prime} \to 0$ provided that $S_0 > I(X;V_0|U)$. Next,
\begin{equation}
\begin{aligned}
& P^{\prime\prime} \triangleq \Pr \left\{\exists \hat{s}_0\in [1:\exp{(nS_0)}] \right.\\& \qquad\text{ s.t. } \,\vct{V}^n_0(\hat{s}_0) \in \cT_{[V_0|UY]\delta}^n(\vct{u}^n\vct{y}^n),\\&\left.\qquad\qquad B_0\big(\vct{v}_0^n(s_0)\big) =  B_0\big(\vct{v}_0^n(\hat{s}_0)\big)\right\}
\end{aligned}
\end{equation}
\begin{equation}
\begin{aligned}
&\leq \exp{[n(S_0-R_0+\epsilon)]}\\& \times\Pr\{(\vct{V}_0^n,\vct{Y}^n) \in \cT_{[V_0Y|U]\delta} (\vct{u}^n) | \vct{V}_0^n \in \cT_{[V_0|U]\delta} (\vct{u}^n), \\& \qquad\qquad\qquad\qquad\qquad\qquad\quad \vct{Y}^n \in \cT_{[Y|U]\delta} (\vct{u}^n)\}\\
&\leq \exp{[n(S_0-R_0+\epsilon)]}\exp{\left[-n\big(Y;V_0|U)+\eta_n^{(2)}\big)\right]}\\
&=\exp{\left\{-n\big[I(Y;V_0|U)-(S_0-R_0)+\eta_n^{(2)}-\epsilon\big]\right\}}\ .
\end{aligned}
\end{equation}
Here, $B_0(\vct{v}_0^n(s_0))$ denotes the bin $\vct{v}_0^n(s_0)$ belongs to, as defined as part of the encoding strategy. $R_0$ is the rate dedicated to the estimation part, for the case that $H_0$ was chosen as the correct hypothesis. Defining $R_1$ equivalently for hypothesis $H_1$, the total available rate can be said to be divided, under the proposed achievable scheme, to three parts, such that $R = R' + R_0 + R_1$. 
Thus, $P^{\prime\prime} \to 0$ if $S_0 - R_0 < I(X;V_0|U)$, or equivalently
\begin{equation}
\begin{aligned}
R_0 &> S_0 - I(Y;V_0|U) \\
&> I(X;V_0U) - I(Y;V_0|U) \\ 
&=I(XY;V_0|U) - I(Y;V_0|U) \\
&= I(X;V_0|UY) \ .
\end{aligned}
\end{equation}
Thus, the probability of error related to source reconstruction goes to zero provided that $S_0 > I(X;V_0|U)$ and $R_0 > I(X;V_0|UY)$. Combining this result with the symmetric case of $H_1$ and the result for the detection step, the required total rate of communication reads
\begin{equation}
R> R^{\prime} + I(X;V_0|UY) + I(\bX;V_1|\bU\bY) \ .
\end{equation} 

We now know that our scheme allows the decoding of either $\vct{v}_0$ and $\vct{v}_1$, depending on the case, with high probability, when $n \to \infty$. It remains to be shown that using the sequence $\vct{v}_0^n$, it is possible to recover $\vct{x}^n$ with distortion $D_0$. We choose a (possibly suboptimal) decoder, that reconstructs $\vct{x}^n$ only from $(\vct{u}^n,\vct{y}^n,\vct{v}_{0}^n)$:
\begin{equation}
\begin{aligned}
& d(\vct{x}^n,\vct{\hat{x}}^n(\vct{u}^n,\vct{y}^n,\vct{v}^n_0)) = \frac{1}{n} \sum\limits_{i=1}^n d\big(x_i,\hat{x}_i(u^n,y^n,v_{0}^n)\big)\\
& = \frac{1}{n} \sum %\limits_{\forall (x,u,y,v_{0})} 
d\big(x,\hat{x}(u,y,v_0)\big)N(x,u,y,v_0|\vct{x}^n\vct{u}^n\vct{y}^n\vct{v}^n_0)\\
&\leq \mathbb{E}_0\left[d\big(X,\hat{X}(UYV_0)\big) \right] \\&+ \sum\left|\frac{1}{n}N(x,u,y,v_0|\vct{x}^n \vct{u}^n\vct{y}^n\vct{v}^n_0) - p(x,u,y,v_0)\right|\\
&\leq \mathbb{E}_0\left[d\big(X,\hat{X}(UYV_0)\big) \right] + d_{\text{max}}|\mathcal{X}||\mathcal{Y}||\mathcal{U}||\mathcal{V}_0|\delta_n \ ,
\end{aligned}
\end{equation}
where the summation is over all the possible letters in the respective alphabets of the RVs, and the final inequality holds since $(\vct{x}^n,\vct{y}^n,\vct{u}^n,\vct{v}^n_0) \in \cT_{[XYUV_0]\delta}^n$. Since $\delta_n \to 0$ when $n \to \infty$, any distortion $D_0$ can be achieved, as long as $D_0 > \mathbb{E}_0\left[d\big(X,\hat{X}(UYV_0)\big)\right]$.

\section{Proof of Proposition~\ref{Prop:BetterAchievable}}\label{Apen:BetterAchievability}
We now prove the achievability of the error exponent offered in Proposition~\ref{Prop:BetterAchievable}, for the case where source reconstruction is not required. As the proof is in many ways similar to the proof of Proposition~\ref{Prop:General}, given in Appendix~\ref{Appen:GeneralAchievability}, we concentrate mainly on the main differences. %Specifically, as the codebook generation and coding scheme are similar to the previous case, we specify only the decoding scheme, before calculating the probability of error.

\subsubsection*{Codebook generation and encoding strategy}
Both the codebook generation and the encoding strategy in this case are very similar to what was done in the proof of Proposition~\ref{Prop:General}, in the part dedicated to detection. The only difference is that now we choose to only work with $\delta$-typical sequences, for some arbitrary $\delta$. When node $A$ sees a non-typical sequence $\vct{x}$, it sends an error message. In the opposite case, encoding is done as before. Note that while we only work with $\delta$-typical sequences, there are still different codebooks for each type \emph{within the set} of $\delta$-typical sequences. 

\subsubsection*{Decoding strategy}
In case an error message is received, the decoder declares  $H_1$. This strategy implies that any probability of the error event caused by the encoder not seeing a $\delta$-typical sequence is  allocated to $\alpha_n$, rather than $\beta_n$. The probability of this event, however, goes to zero when $n \to \infty$ thanks to the AEP, implying that $\alpha_n \leq \epsilon$ for any $\epsilon >0$, for $n \geq n_0(\epsilon,\delta)$, thus satisfying the constraint over $\alpha_n$.

When the encoder does not send an error message, the decoder operates on the entire bin in order to make a decision. Going over the sequences in the bin one by one, the decoder checks for each $\vct{u}^n_i$ if $(\vct{u}^n_i,\vct{y}^n) \in T_{[UY]\delta}^n$. If a sequence in the bin is found, which is jointly typical with $\vct{y}^n$, the decoder declares $H_0$. If no such sequence is found, the decoder declares $H_1$. Note that under this strategy, the decoder does not attempt to find the original sequence sent by the encoder. Specifically, when the decoder declares $H_1$ it is completely oblivious to the original codeword.

\subsubsection*{Probability of error}
The analysis of the probability of error in detection under this new strategy is very similar to the analysis given in Appendix~\ref{Appen:GeneralAchievability}. We  separately bound the corresponding error probabilities on the two possible error events.

\emph{Analysis of $\alpha_n$:} When analyzing $\alpha_n(\mathcal{A}_n)=\Pr\big(\mathcal{A}_n^c | XY \sim p_0 (x,y) \big)$, we assume throughout that the  probability measure in effect is $p_0$. Two scenarios can lead to an event where the decoder erroneously declares $H_1$:
\begin{equation}
\begin{aligned}
&\mathcal{B}_{13} \triangleq  \big\{\nexists\, i \in \mathcal{C}_U^n(Q_{\vct{x}^n}) \,\big|\, (\vct{x}^n,\vct{u}^n_i) \in \cT_{[UX]\delta}^n\big\}\ ,\\
&\mathcal{B}_{14} \triangleq \big\{\nexists \,i \in F(f(\vct{x}^n)) \,\big|\,  (\vct{u}_i^n,\vct{y}^n) \in \cT_{[UY]\delta}^n\big\} \ .
\end{aligned}
\end{equation}
In the first event, an error message is sent, as there is no fitting codeword within  the codebook for the observed sequence $\vct{x}^n$. Whereas for the second event, there is no sequence in the bin that prompts the decoder to decide $H_0$, despite it being the true hypothesis. 
The probability of event $\mathcal{B}_{13}$ goes to zero with $n$, thanks to the AEP and the size of the codebook. As for event $\mathcal{B}_{14}$, assume without loss of generality, that the encoder intended to send the first word in the bin $\vct{u}_1^n$, i.e., $\vct{u}^n_1 = f(\vct{x}^n)$. The probability that the decoder declares $H_1$ can be upper-bounded by 
\begin{equation}
\begin{aligned}
\Pr (\cB_{14}) & = \Pr\big\{\nexists \,i \in F(f(\vct{X}^n)) \,\big|\,  (\vct{U}_i^n,\vct{Y}^n) \in \cT_{[UY]\delta}^n\big\}\\&\quad\leq  \Pr\{ (\vct{U}_1^n,\vct{Y}^n)  \notin \cT_{[UY]\delta}^n \} \ ,
\end{aligned}
\end{equation}
where typicality is measured over the probability measure  $p_{0} = P_{XY}$. As was already discussed above, this probability tends to $0$ with the number of available realizations $n$. This result is attributed to the AEP, by which $\vct{x}$ and $\vct{y}$ are jointly typical with high probability, and to the generalized Markov Lemma (Lemma~\ref{lemma:markov}). Thus, any fixed constraint over the probability of error of the first type $\alpha \leq \epsilon$  ($\epsilon>0$), may be satisfied when $n$ is large enough.

\emph{Analysis of $\beta_n$:} As we now turn to analyzing the probability of error of the second type, we assume throughout this part that the real hypothesis is $H_1$. As was the case in Appendix~\ref{Appen:GeneralAchievability}, the resulting error exponent is the result of a trade-off between two error events. While the analysis of the event where the correct sequence prompts a wrong decision (i.e. in this case is $(f(\vct{x}^n),\vct{y}^n) \in \cT_{[UY]\delta}^n$) stays the same, the second error event is now different. We thus concentrate in this appendix on calculating the probability of the event that some sequence in the bin $\vct{u}^n \neq f(\vct{x}^n)$ prompts the decoder to declare $H_0$. We start by presenting the following lemma:
\begin{lemma}\label{LemmaBinningNew} Let $\mathcal{A}_n$ be the set of triplets, such that a binned codebook is necessary:
	\begin{equation}
	\begin{aligned}
		\mathcal{A}_n = \Big\{&(\vct{u}^n,\vct{x}^n,\vct{y}^n)\in T_{Q_{{U}|{X}}^\star}^n \times \mathcal{X}^n\times \mathcal{Y}^n\big| \\ & 
			\qquad\qquad\log|\mathcal{C}_U^n(Q_{\vct{x}^n})| \geq nR\Big\} \ .
	\end{aligned}
	\end{equation}
	Let $(\vct{u}^n,\vct{x}^n,\vct{y}^n) \in \mathcal{A}_n$ and denote by $\mathcal{B}_{15}$ the event indicating that  $(\vct{u}^n,\vct{y}^n) \in \cT_{[UY]\delta}^n$, for some $\vct{u}^n \neq f(\vct{x}^n)$ in the bin. 
	Then,
	\begin{equation}\label{Eq:LemmaProbability2}
	\begin{aligned}
	& \Pr \left(\mathcal{B}_{15}|\vct{U}^n=\vct{u}^n,\vct{X}^n = \vct{x}^n, \vct{Y}^n=\vct{y}^n\right) \\&\leq \exp\left[{-n\left(R-\hat{J}(Q_{{\vct{u}^n}{\vct{x}^n}{\vct{y}^n}})-\delta_n\right)}\right] \ ,
	\end{aligned}
	\end{equation} 
	with
	\begin{equation}
	\begin{aligned}
	\hat{J}(Q_{{\vct{u}^n}{\vct{x}^n}{\vct{y}^n}}) \triangleq & I\big(Q_{{\vct{x}^n}};Q_{U|X}^\star\big) - H(Q_{{\vct{u}^n}}) \\&+ H\big(Q_{U|Y}|P_{Y}\big)
	\end{aligned}
	\end{equation}
	and
	\begin{equation}
	\delta_n \triangleq \frac{1}{n} \log(n+1)^{|\mathcal{U}|(1+|\mathcal{X}|+|\mathcal{Y}|) +4} + \epsilon_n \ 
	\end{equation}
	with $\epsilon_n \to 0 $ when $n \to \infty$. Moreover, the probability in \eqref{Eq:LemmaProbability2} is taken over the choice of the codebook in use.
\end{lemma}

\begin{IEEEproof}
	The proof of Lemma~\ref{LemmaBinningNew} is very similar to the one given for Lemma~\ref{LemmaBinning}. The difference is that now the set of sequences that ``confuses'' the decoder is simply $\hat{\mathcal{S}}(\vct{y}^n)=\cT_{[U|Y]\delta}^n(\vct{y}^n)$. Bounding the set of conditionally typical sequences by~\cite{ElGamal-Kim-2011}:
	\begin{equation}
	\begin{aligned}
		& \left| \cT_{[U|Y]\delta}^n(\vct{y}^n) \right| \\& \quad\leq (n+1)^{|\mathcal{U}| |\mathcal{Y}|} \exp\big[{n(H(Q_{U|Y}|P_{Y}) + \epsilon_n)}\big]  \ ,
	\end{aligned}
	\end{equation}
	for each $\vct{y}^n\in \cT_{[Y]\delta}^n$, completes the proof.
\end{IEEEproof}
\begin{Remark}
	Note that unlike $J(Q_{{\vct{u}^n}{\vct{x}^n}{\vct{y}^n}})$, the quantity $\hat{J}(Q_{{\vct{u}^n}{\vct{x}^n}{\vct{y}^n}})$ is not dependent on the observed $\vct{y}^n$. The quantity $H(Q_{U|Y}|P_{Y})$ can be analytically calculated when the type of $\vct{x}^n$ and the chosen strategy $Q_{U|X}$ is known, without knowing neither the specific sent sequence $\vct{u}^n$ nor the observed sequence $\vct{y}^n$.
\end{Remark}

Using Lemma~\ref{LemmaBinningNew} and summing over all involved types and sequences within each type as was done in Appendix~\ref{Appen:GeneralAchievability}, the probability of the event where an unintended sequence in the bin causes an error can be bounded by
\begin{equation}
\begin{aligned}
& \lim\limits_{n \to \infty} -\frac{1}{n} \log \Pr(\mathcal{B}_{15}) \geq \nonumber \\
&\min\limits_{Q_X\in \mathcal{P}_n(\mathcal{X})}\, \max\limits_{Q_{U|X}^\star(Q_X) \in \mathcal{P}_n(\mathcal{U})} \, \min\limits_{Q_Y\in \mathcal{P}_n(\mathcal{Y})}\, \min\limits_{Q_{UXY}\in \mathcal{P}_n(\mathcal{U}\times \mathcal{X} \times \mathcal{Y}) }\\&\qquad\Big\{\mathcal{D}(Q_{UXY}\|P_{\bU\bX\bY}) + R - \hat{J}(Q_{UXY})\Big\}\nonumber \\
&=\min\limits_{Q_X}\, \max\limits_{Q_{U|X}^\star(Q_X) } \, \min\limits_{Q_Y}\, \min\limits_{Q_{UXY}} \\&\qquad\Big \{\mathcal{D}(Q_{UXY}\|P_{\bU\bX\bY})  +R \nonumber \\
&\qquad - I(Q_X;Q_{U|X}^\star) + I(Q_{U|Y}^\star;P_{Y}) \Big\} \ .\label{eq:minmax}
\end{aligned}
\end{equation}
As in this case we only work with $\delta$-typical ${x}$-sequences, we may choose $\delta$ to be any value, as long as it is strictly positive. Thus, we may force $Q_X$ to be arbitrarily close to $P_X$ by taking $\delta \to 0^+$. The error exponent in question thus becomes 
\begin{equation}
\begin{aligned}
\lim\limits_{n \to \infty} & -\frac{1}{n} \log \Pr(\mathcal{B}_{15})\nonumber\\
&\geq\max\limits_{Q_{U|X}^\star\in \mathcal{P}(\mathcal{U})} \Big\{R - I(P_X;Q_{U|X}^\star) \nonumber + I(P_{Y};Q_{U|Y}^\star) \\&+ \min\limits_{Q_Y\in \mathcal{P}(\mathcal{Y})}\min\limits_{Q_{UXY}\in \mathcal{P}(\mathcal{U}\times\mathcal{X}\times\mathcal{Y})} \mathcal{D}(Q_{UXY}\|P_{\bU\bX\bY}) \Big\} + \hat{\epsilon} \\
& = \max\limits_{Q_{U|X}^\star\in \mathcal{P}(\mathcal{U})} \Big\{R - I(P_X;Q_{U|X}^\star) + I(P_{Y};Q_{U|Y}^\star) \Big\} + \hat{\epsilon} \ ,
\end{aligned}
\end{equation}
with $\hat{\epsilon} \to 0$ as $\delta \to 0$. This, along with an analysis of the complementary error event similar to the one given for Proposition~\ref{Prop:General}, completes the  proof of Proposition~\ref{Prop:BetterAchievable}.

%\textcolor{red}{\# PLEASE CHECK THE PREVIOUS STEPS, WHAT ABOUT CONTINUITY ASSUMPTIONS? IS THE LAST EXPRESSION A MINIMIZATION OVER TYPES OR PROBABILITIES?} \textcolor{blue}{As in this expression we take the limit over $n$ from the beginning (and do not bother calculating for finite $n$ and then taking the limit), these are probabilities. I replaced all $\mathcal{P}_n$ with $\mathcal{P}$.}
\section*{Acknowledgment}

The authors are grateful to Prof.\ Romain Couillet for his valuable comments at the early stage of this work. They are also grateful to the Associate Editor, and to anonymous reviewers for their constructive and helpful comments on the earlier version of the manuscript.

\bibliographystyle{IEEEtran.bst}
\bibliography{IEEEabrv,HypothesisTestingBib}

\end{document}

%% file: figures/error_distortion_independence.tex
\begin{tikzpicture}
	\begin{axis}[
	title = \textbf{},
		xlabel=$E$,
		ylabel=$D\text{[BER]}$,
		xmin = 0,
		ymin = 0,
		xmax = 0.1887,
		ymax = 0.25]
		\pgfplotsset{every axis legend/.append style={
at={(0.98,0.3),font = \small},
anchor= south east}}

%	\addplot[blue,line width=1pt] coordinates {
%(0.500,0.681) (0.600,0.615) (0.700,0.566) (0.800,0.529) (0.900,0.499) (1.000,0.475) (1.100,0.455) (1.200,0.438) (1.300,0.424) (1.400,0.411) (1.500,0.401) (1.600,0.391) (1.700,0.383) (1.800,0.375) (1.900,0.368) (2.000,0.362) (2.100,0.357) (2.200,0.352) (2.300,0.347) (2.400,0.343) 
%	};
%	
%	\addplot[red,line width=1pt] coordinates {
%(0.500,0.613) (0.600,0.543) (0.700,0.491) (0.800,0.452) (0.900,0.421) (1.000,0.396) (1.100,0.375) (1.200,0.358) (1.300,0.343) (1.400,0.330) (1.500,0.319) (1.600,0.310) (1.700,0.301) (1.800,0.294) (1.900,0.287) (2.000,0.281) (2.100,0.275) (2.200,0.270) (2.300,0.266) (2.400,0.261) 
%	};
	
%	\addplot[black,dotted,line width=1pt] coordinates {
%(1.100,0.346) (1.200,0.329) (1.300,0.314) (1.400,0.301) (1.500,0.289) (1.600,0.278) (1.700,0.268) (1.800,0.260) (1.900,0.252) (2.000,0.244) (2.100,0.238) (2.200,0.231) (2.300,0.226) (2.400,0.220) (2.500,0.215) (2.600,0.211) (2.700,0.206) (2.800,0.202) (2.900,0.199) (3.000,0.195) 
%	};

\addplot[black,line width = 1pt] coordinates{
( 0.000, 0.214)( 0.010, 0.217)( 0.020, 0.221)( 0.030, 0.224)( 0.040, 0.228)( 0.050, 0.232)( 0.060, 0.235)( 0.070, 0.239)( 0.080, 0.243)( 0.090, 0.247)( 0.100, 0.250)
};
	
\addplot[blue,line width = 1pt] coordinates{
( 0.000, 0.177)( 0.019, 0.184)( 0.038, 0.191)( 0.057, 0.198)( 0.075, 0.205)( 0.094, 0.211)( 0.113, 0.218)( 0.132, 0.225)( 0.151, 0.232)( 0.170, 0.239)( 0.189, 0.246)
};

\addplot[red,line width = 1pt] coordinates{
( 0.000, 0.069)( 0.019, 0.075)( 0.038, 0.082)( 0.057, 0.088)( 0.075, 0.095)( 0.094, 0.102)( 0.113, 0.109)( 0.132, 0.116)( 0.151, 0.123)( 0.170, 0.130)( 0.189, 0.136)
};

\addplot[green,line width = 1pt] coordinates{
( 0.000, 0.001)( 0.019, 0.006)( 0.038, 0.008)( 0.057, 0.011)( 0.075, 0.014)( 0.094, 0.017)( 0.113, 0.021)( 0.132, 0.025)( 0.151, 0.029)( 0.170, 0.033)( 0.189, 0.037)
};

\addplot[orange,line width = 1pt] coordinates{
( 0.000, 0.000)( 0.019, 0.000)( 0.038, 0.000)( 0.057, 0.000)( 0.075, 0.000)( 0.094, 0.000)( 0.113, 0.005)( 0.132, 0.007)( 0.151, 0.010)( 0.170, 0.013)( 0.189, 0.016)
};

\addplot[purple,line width = 1pt] coordinates{
( 0.000, 0.000)( 0.019, 0.000)( 0.038, 0.000)( 0.057, 0.000)( 0.075, 0.000)( 0.094, 0.000)( 0.113, 0.000)( 0.132, 0.000)( 0.151, 0.000)( 0.170, 0.000)( 0.189, 0.000)

};

	\legend{ {$R=0.1$},{$R=0.2$},{$R=0.5$}, {$R=0.8$},{$R=0.9$},{$R=1$}}
	\end{axis}

\end{tikzpicture}

%% file: figures/BSCGeneral.tex
\begin{tikzpicture}
	\begin{axis}[
	title = \textbf{},
		xlabel=$\delta$,
		ylabel=,
		xmin = 0,
		ymin = 0,
		xmax = 0.18,
		ymax = 0.2]
		\pgfplotsset{every axis legend/.append style={
at={(0.02,0.98),font = \small},
anchor= north west}}

%	\addplot[blue,line width=1pt] coordinates {
%(0.500,0.681) (0.600,0.615) (0.700,0.566) (0.800,0.529) (0.900,0.499) (1.000,0.475) (1.100,0.455) (1.200,0.438) (1.300,0.424) (1.400,0.411) (1.500,0.401) (1.600,0.391) (1.700,0.383) (1.800,0.375) (1.900,0.368) (2.000,0.362) (2.100,0.357) (2.200,0.352) (2.300,0.347) (2.400,0.343) 
%	};
%	
%	\addplot[red,line width=1pt] coordinates {
%(0.500,0.613) (0.600,0.543) (0.700,0.491) (0.800,0.452) (0.900,0.421) (1.000,0.396) (1.100,0.375) (1.200,0.358) (1.300,0.343) (1.400,0.330) (1.500,0.319) (1.600,0.310) (1.700,0.301) (1.800,0.294) (1.900,0.287) (2.000,0.281) (2.100,0.275) (2.200,0.270) (2.300,0.266) (2.400,0.261) 
%	};
	
%	\addplot[black,dotted,line width=1pt] coordinates {
%(1.100,0.346) (1.200,0.329) (1.300,0.314) (1.400,0.301) (1.500,0.289) (1.600,0.278) (1.700,0.268) (1.800,0.260) (1.900,0.252) (2.000,0.244) (2.100,0.238) (2.200,0.231) (2.300,0.226) (2.400,0.220) (2.500,0.215) (2.600,0.211) (2.700,0.206) (2.800,0.202) (2.900,0.199) (3.000,0.195) 
%	};

%\draw[fill=gray!50,opacity=.7]
%(axis cs:0,0) -- (axis cs:0.0815,0) -- (axis cs:0.0815,0.2) -- (axis cs:0,0.2) -- cycle ;

\addplot[black,line width = 1pt] coordinates{
( 0.00100503, 0.05253741)( 0.00201005, 0.05214594)( 0.00301508, 0.05175804)( 0.00402010, 0.05137366)( 0.00502513, 0.05099275)( 0.00603015, 0.05061526)( 0.00703518, 0.05024115)( 0.00804020, 0.04987038)( 0.00904523, 0.04950289)( 0.01005025, 0.04913865)( 0.01105528, 0.04877762)( 0.01206030, 0.04841974)( 0.01306533, 0.04806499)( 0.01407035, 0.04771332)( 0.01507538, 0.04736469)( 0.01608040, 0.04701906)( 0.01708543, 0.04667639)( 0.01809045, 0.04633665)( 0.01909548, 0.04599980)( 0.02010050, 0.04566580)( 0.02110553, 0.04533463)( 0.02211055, 0.04500623)( 0.02311558, 0.04468059)( 0.02412060, 0.04435766)( 0.02512563, 0.04403741)( 0.02613065, 0.04371981)( 0.02713568, 0.04340483)( 0.02814070, 0.04309244)( 0.02914573, 0.04278260)( 0.03015075, 0.04247529)( 0.03115578, 0.04217048)( 0.03216080, 0.04186813)( 0.03316583, 0.04156821)( 0.03417085, 0.04127071)( 0.03517588, 0.04097559)( 0.03618090, 0.04068282)( 0.03718593, 0.04039238)( 0.03819095, 0.04010424)( 0.03919598, 0.03981837)( 0.04020101, 0.03953475)( 0.04120603, 0.03925336)( 0.04221106, 0.03897416)( 0.04321608, 0.03869713)( 0.04422111, 0.03842226)( 0.04522613, 0.03814951)( 0.04623116, 0.03787887)( 0.04723618, 0.03761030)( 0.04824121, 0.03734379)( 0.04924623, 0.03707931)( 0.05025126, 0.03681685)( 0.05125628, 0.03655638)( 0.05226131, 0.03629788)( 0.05326633, 0.03604132)( 0.05427136, 0.03578670)( 0.05527638, 0.03553398)( 0.05628141, 0.03528315)( 0.05728643, 0.03503418)( 0.05829146, 0.03478707)( 0.05929648, 0.03454178)( 0.06030151, 0.03429831)( 0.06130653, 0.03405662)( 0.06231156, 0.03381671)( 0.06331658, 0.03357856)( 0.06432161, 0.03334214)( 0.06532663, 0.03310744)( 0.06633166, 0.03287445)( 0.06733668, 0.03264314)( 0.06834171, 0.03241349)( 0.06934673, 0.03218550)( 0.07035176, 0.03195915)( 0.07135678, 0.03173441)( 0.07236181, 0.03151128)( 0.07336683, 0.03128974)( 0.07437186, 0.03106977)( 0.07537688, 0.03085136)( 0.07638191, 0.03063448)( 0.07738693, 0.03041914)( 0.07839196, 0.03020531)( 0.07939698, 0.02999298)( 0.08040201, 0.02978213)( 0.08140704, 0.02957275)( 0.08241206, 0.02936482)( 0.08341709, 0.02915834)( 0.08442211, 0.02895329)( 0.08542714, 0.02874965)( 0.08643216, 0.02854742)( 0.08743719, 0.02834658)( 0.08844221, 0.02814711)( 0.08944724, 0.02794901)( 0.09045226, 0.02775225)( 0.09145729, 0.02755684)( 0.09246231, 0.02736276)( 0.09346734, 0.02716999)( 0.09447236, 0.02697852)( 0.09547739, 0.02678834)( 0.09648241, 0.02659945)( 0.09748744, 0.02641182)( 0.09849246, 0.02622545)( 0.09949749, 0.02604033)( 0.10050251, 0.02585645)( 0.10150754, 0.02567378)( 0.10251256, 0.02549234)( 0.10351759, 0.02531209)( 0.10452261, 0.02513304)( 0.10552764, 0.02495517)( 0.10653266, 0.02477847)( 0.10753769, 0.02460294)( 0.10854271, 0.02442856)( 0.10954774, 0.02425532)( 0.11055276, 0.02408322)( 0.11155779, 0.02391223)( 0.11256281, 0.02374237)( 0.11356784, 0.02357360)( 0.11457286, 0.02340594)( 0.11557789, 0.02323936)( 0.11658291, 0.02307385)( 0.11758794, 0.02290942)( 0.11859296, 0.02274604)( 0.11959799, 0.02258372)( 0.12060302, 0.02242244)( 0.12160804, 0.02226219)( 0.12261307, 0.02210297)( 0.12361809, 0.02194476)( 0.12462312, 0.02178757)( 0.12562814, 0.02163138)( 0.12663317, 0.02147618)( 0.12763819, 0.02132196)( 0.12864322, 0.02116872)( 0.12964824, 0.02101645)( 0.13065327, 0.02086515)( 0.13165829, 0.02071480)( 0.13266332, 0.02056539)( 0.13366834, 0.02041693)( 0.13467337, 0.02026939)( 0.13567839, 0.02012279)( 0.13668342, 0.01997710)( 0.13768844, 0.01983232)( 0.13869347, 0.01968844)( 0.13969849, 0.01954547)( 0.14070352, 0.01940338)( 0.14170854, 0.01926218)( 0.14271357, 0.01912185)( 0.14371859, 0.01898240)( 0.14472362, 0.01884380)( 0.14572864, 0.01870607)( 0.14673367, 0.01856919)( 0.14773869, 0.01843315)( 0.14874372, 0.01829795)( 0.14974874, 0.01816358)( 0.15075377, 0.01803004)( 0.15175879, 0.01789731)( 0.15276382, 0.01776541)( 0.15376884, 0.01763431)( 0.15477387, 0.01750401)( 0.15577889, 0.01737451)( 0.15678392, 0.01724580)( 0.15778894, 0.01711788)( 0.15879397, 0.01699073)( 0.15979899, 0.01686436)( 0.16080402, 0.01673876)( 0.16180905, 0.01661392)( 0.16281407, 0.01648984)( 0.16381910, 0.01636651)( 0.16482412, 0.01624393)( 0.16582915, 0.01612209)( 0.16683417, 0.01600099)( 0.16783920, 0.01588062)( 0.16884422, 0.01576098)( 0.16984925, 0.01564206)( 0.17085427, 0.01552385)( 0.17185930, 0.01540636)( 0.17286432, 0.01528958)( 0.17386935, 0.01517349)( 0.17487437, 0.01505811)( 0.17587940, 0.01494342)( 0.17688442, 0.01482941)( 0.17788945, 0.01471609)( 0.17889447, 0.01460345)( 0.17989950, 0.01449149)( 0.18090452, 0.01438019)( 0.18190955, 0.01426956)( 0.18291457, 0.01415959)( 0.18391960, 0.01405028)( 0.18492462, 0.01394162)( 0.18592965, 0.01383361)( 0.18693467, 0.01372625)( 0.18793970, 0.01361952)( 0.18894472, 0.01351344)( 0.18994975, 0.01340798)( 0.19095477, 0.01330315)( 0.19195980, 0.01319895)( 0.19296482, 0.01309536)( 0.19396985, 0.01299239)( 0.19497487, 0.01289004)( 0.19597990, 0.01278829)( 0.19698492, 0.01268715)( 0.19798995, 0.01258661)( 0.19899497, 0.01248667)
};
	
\addplot[blue,line width = 1pt] coordinates{
( 0.00100503, 0.00001007)( 0.07738693, 0.00011575)( 0.08140704, 0.00040739)( 0.09246231, 0.00166696)( 0.10150754, 0.00511033)( 0.10954774, 0.01165135)( 0.11356784, 0.01576694)( 0.11959799, 0.02279674)( 0.12864322, 0.03945510)( 0.19798995, 0.17617448)( 0.19899497, 0.17869036)

};

\addplot[red,line width = 1pt,dashed] coordinates{
( 0.004, 0.019)( 0.008, 0.019)( 0.012, 0.019)( 0.016, 0.019)( 0.020, 0.019)( 0.024, 0.019)( 0.029, 0.019)( 0.033, 0.019)( 0.037, 0.019)( 0.041, 0.019)( 0.045, 0.019)( 0.049, 0.019)( 0.053, 0.019)( 0.057, 0.019)( 0.061, 0.019)( 0.065, 0.019)( 0.069, 0.019)( 0.073, 0.019)( 0.078, 0.019)( 0.082, 0.019)( 0.086, 0.019)( 0.090, 0.019)( 0.094, 0.019)( 0.098, 0.019)( 0.102, 0.019)( 0.106, 0.019)( 0.110, 0.019)( 0.114, 0.019)( 0.118, 0.019)( 0.122, 0.019)( 0.127, 0.019)( 0.131, 0.019)( 0.135, 0.019)( 0.139, 0.019)( 0.143, 0.019)( 0.147, 0.019)( 0.151, 0.019)( 0.155, 0.019)( 0.159, 0.019)( 0.163, 0.019)( 0.167, 0.019)( 0.171, 0.019)( 0.176, 0.019)( 0.180, 0.019)( 0.184, 0.019)( 0.188, 0.019)( 0.192, 0.019)( 0.196, 0.019)

};

\addplot[green,line width = 1pt] coordinates{
( 0.001, 0.000)( 0.002, 0.000)( 0.003, 0.000)( 0.004, 0.000)( 0.005, 0.000)( 0.006, 0.000)( 0.007, 0.000)( 0.008, 0.000)( 0.009, 0.000)( 0.010, 0.000)( 0.011, 0.000)( 0.012, 0.000)( 0.013, 0.000)( 0.014, 0.003)( 0.015, 0.007)( 0.016, 0.010)( 0.017, 0.014)( 0.018, 0.017)( 0.019, 0.021)( 0.020, 0.024)( 0.021, 0.027)( 0.022, 0.030)( 0.023, 0.034)( 0.024, 0.037)( 0.025, 0.040)( 0.026, 0.043)( 0.027, 0.046)( 0.028, 0.048)( 0.029, 0.051)( 0.030, 0.054)( 0.031, 0.057)( 0.032, 0.060)( 0.033, 0.062)( 0.034, 0.065)( 0.035, 0.067)( 0.036, 0.070)( 0.037, 0.072)( 0.038, 0.075)( 0.039, 0.077)( 0.040, 0.080)( 0.041, 0.082)( 0.042, 0.085)( 0.043, 0.087)( 0.044, 0.089)( 0.045, 0.092)( 0.046, 0.094)( 0.047, 0.096)( 0.048, 0.098)( 0.049, 0.101)( 0.050, 0.103)( 0.051, 0.105)( 0.052, 0.107)( 0.053, 0.109)( 0.054, 0.111)( 0.055, 0.113)( 0.056, 0.115)( 0.057, 0.117)( 0.058, 0.119)( 0.059, 0.121)( 0.060, 0.123)( 0.061, 0.125)( 0.062, 0.127)( 0.063, 0.129)( 0.064, 0.131)( 0.065, 0.133)( 0.066, 0.135)( 0.067, 0.136)( 0.068, 0.138)( 0.069, 0.140)( 0.070, 0.142)( 0.071, 0.144)( 0.072, 0.145)( 0.073, 0.147)( 0.074, 0.149)( 0.075, 0.151)( 0.076, 0.152)( 0.077, 0.154)( 0.078, 0.156)( 0.079, 0.157)( 0.080, 0.159)( 0.081, 0.161)( 0.082, 0.162)( 0.083, 0.164)( 0.084, 0.166)( 0.085, 0.167)( 0.086, 0.169)( 0.087, 0.170)( 0.088, 0.172)( 0.089, 0.173)( 0.090, 0.175)( 0.091, 0.176)( 0.092, 0.178)( 0.093, 0.179)( 0.094, 0.181)( 0.095, 0.182)( 0.096, 0.184)( 0.097, 0.185)( 0.098, 0.187)( 0.099, 0.188)( 0.101, 0.190)( 0.102, 0.191)( 0.103, 0.192)( 0.104, 0.194)( 0.105, 0.195)( 0.106, 0.197)( 0.107, 0.198)( 0.108, 0.199)( 0.109, 0.201)( 0.110, 0.202)( 0.111, 0.203)( 0.112, 0.205)( 0.113, 0.206)( 0.114, 0.207)( 0.115, 0.209)( 0.116, 0.210)( 0.117, 0.211)( 0.118, 0.212)( 0.119, 0.214)( 0.120, 0.215)( 0.121, 0.216)( 0.122, 0.217)( 0.123, 0.219)( 0.124, 0.220)( 0.125, 0.221)( 0.126, 0.222)( 0.127, 0.224)( 0.128, 0.225)( 0.129, 0.226)( 0.130, 0.227)( 0.131, 0.228)( 0.132, 0.230)( 0.133, 0.231)( 0.134, 0.232)( 0.135, 0.233)( 0.136, 0.234)( 0.137, 0.235)( 0.138, 0.236)( 0.139, 0.238)( 0.140, 0.239)( 0.141, 0.240)( 0.142, 0.241)( 0.143, 0.242)( 0.144, 0.243)( 0.145, 0.244)( 0.146, 0.245)( 0.147, 0.246)( 0.148, 0.247)( 0.149, 0.248)( 0.150, 0.249)( 0.151, 0.250)( 0.152, 0.251)( 0.153, 0.253)( 0.154, 0.254)( 0.155, 0.255)( 0.156, 0.256)( 0.157, 0.257)( 0.158, 0.258)( 0.159, 0.259)( 0.160, 0.260)( 0.161, 0.261)( 0.162, 0.262)( 0.163, 0.263)( 0.164, 0.264)( 0.165, 0.264)( 0.166, 0.265)( 0.167, 0.266)( 0.168, 0.267)( 0.169, 0.268)( 0.170, 0.269)( 0.171, 0.270)( 0.172, 0.271)( 0.173, 0.272)( 0.174, 0.273)( 0.175, 0.274)( 0.176, 0.275)( 0.177, 0.276)( 0.178, 0.277)( 0.179, 0.277)( 0.180, 0.278)( 0.181, 0.279)( 0.182, 0.280)( 0.183, 0.281)( 0.184, 0.282)( 0.185, 0.283)( 0.186, 0.284)( 0.187, 0.284)( 0.188, 0.285)( 0.189, 0.286)( 0.190, 0.287)( 0.191, 0.288)( 0.192, 0.289)( 0.193, 0.289)( 0.194, 0.290)( 0.195, 0.291)( 0.196, 0.292)( 0.197, 0.293)( 0.198, 0.294)( 0.199, 0.294)
};

\addplot[purple,line width = 1pt,dashed] coordinates{
	( 0.004, 0.0529)( 0.008, 0.0529)( 0.012, 0.0529)( 0.016, 0.0529)( 0.020, 0.0529)( 0.024, 0.0529)( 0.029, 0.0529)( 0.033, 0.0529)( 0.037, 0.0529)( 0.041, 0.0529)( 0.045, 0.0529)( 0.049, 0.0529)( 0.053, 0.0529)( 0.057, 0.0529)( 0.061, 0.0529)( 0.065, 0.0529)( 0.069, 0.0529)( 0.073, 0.0529)( 0.078, 0.0529)( 0.082, 0.0529)( 0.086, 0.0529)( 0.090, 0.0529)( 0.094, 0.0529)( 0.098, 0.0529)( 0.102, 0.0529)( 0.106, 0.0529)( 0.110, 0.0529)( 0.114, 0.0529)( 0.118, 0.0529)( 0.122, 0.0529)( 0.127, 0.0529)( 0.131, 0.0529)( 0.135, 0.0529)( 0.139, 0.0529)( 0.143, 0.0529)( 0.147, 0.0529)( 0.151, 0.0529)( 0.155, 0.0529)( 0.159, 0.0529)( 0.163, 0.0529)( 0.167, 0.0529)( 0.171, 0.0529)( 0.176, 0.0529)( 0.180, 0.0529)( 0.184, 0.0529)( 0.188, 0.0529)( 0.192, 0.0529)( 0.196, 0.0529)
	
};

\addplot[mark=*] coordinates {( 0.11959799, 0.02279674)};
\addplot[mark=*] coordinates {( 0.026, 0.043)};

%\addplot[green,line width = 1pt,dashed] coordinates{
%( 0.004, 0.029)( 0.008, 0.029)( 0.012, 0.029)( 0.016, 0.029)( 0.020, 0.029)( 0.024, 0.029)( 0.029, 0.029)( 0.033, 0.029)( 0.037, 0.029)( 0.041, 0.029)( 0.045, 0.029)( 0.049, 0.029)( 0.053, 0.029)( 0.057, 0.029)( 0.061, 0.029)( 0.065, 0.029)( 0.069, 0.029)( 0.073, 0.029)( 0.078, 0.029)( 0.082, 0.029)( 0.086, 0.029)( 0.090, 0.029)( 0.094, 0.029)( 0.098, 0.029)( 0.102, 0.029)( 0.106, 0.029)( 0.110, 0.029)( 0.114, 0.029)( 0.118, 0.029)( 0.122, 0.029)( 0.127, 0.029)( 0.131, 0.029)( 0.135, 0.029)( 0.139, 0.029)( 0.143, 0.029)( 0.147, 0.029)( 0.151, 0.029)( 0.155, 0.029)( 0.159, 0.029)( 0.163, 0.029)( 0.167, 0.029)( 0.171, 0.029)( 0.176, 0.029)( 0.180, 0.029)( 0.184, 0.029)( 0.188, 0.029)( 0.192, 0.029)( 0.196, 0.029)
%};

	\legend{ {hypothesis testing error exponent},{$G$ (Proposition~\ref{Prop:General})},{hypothesis testing without binning},{$\hat{G}$ (Proposition~\ref{Prop:BetterAchievable})},{``Stein'' Upper Bound}}
	\end{axis}

\end{tikzpicture}